 \documentclass[preprint]{aastex6}

\usepackage{mathtools}
\usepackage{epstopdf}
\usepackage{lineno}

\fullcollaborationName{The Friends of AASTeX Collaboration}

\begin{document}


\title{Photometric Properties of Network and Faculae \\Derived from HMI Data Compensated for Scattered-Light}

\author{Serena Criscuoli\altaffilmark{1},  Aimee Norton\altaffilmark{2}, Taylor Whitney\altaffilmark{1,3}}
\affil{National Solar Observatory, 3665 Discovery Dr., Boulder, CO 80303, USA\\
Hansen Experimental Physics Laboratory, Stanford University, Stanford, CA, 94305 USA\\
University of Nebraska-Lincoln, 1400 R St, Lincoln, NE 68588}
 
\begin{abstract}
We report on the photometric properties of faculae and network as observed in
full-disk, scattered-light corrected images from the Helioseismic Magnetic Imager. We use a Lucy-Richardson deconvolution routine that corrects an 
image in less than one second. Faculae are distinguished from network through proximity to active regions. 
This is the first report that full-disk observations, including center-to-limb variations, reproduce the photometric properties of faculae and network 
observed previously only in sub-arcsecond resolution, small field-of-view studies, i.e. that network, as defined by distance from active regions,
exhibit higher photometric contrasts. 
Specifically, for magnetic flux values larger than approximately 300 G, the network is brighter than faculae and the contrast differences 
increases toward the limb, where the network contrast 
is about twice the facular one. For lower magnetic flux values, network appear darker than faculae. 
Contrary to reports from previous full-disk observations, we also found that network exhibits a higher center-to-limb variation.  
Our results are in agreement with reports from simulations that indicate magnetic flux alone is a poor proxy of the photometric properties of magnetic features.
We estimate that the contribution of faculae and network to Total Solar Irradiance variability of the current 
Cycle 24 is overestimated by at least 11\% due to the photometric properties of network and faculae not being recognized as different. 
This estimate is specific to the method employed in this study to reconstruct irradiance variations, so caution should be paid when extending it to other techniques.     

\end{abstract}

\section{Introduction} \label{sec:intro}
Over the past few decades, much effort has been dedicated to measuring solar irradiance variations and understanding and modeling the physical processes that 
drive them. Motivating this research is the impact that irradiance variations have on the Earth's
atmosphere and climate, especially at the eleven-years solar cycle and longer time scale  \citep[e.g.][]{ermolli2013,seppala2014}.

Variations of solar irradiance at temporal scales longer than one day are modulated by the area and position occupied  over the disk by photospheric magnetic structures. 
Accordingly, various techniques have been developed to reproduce variations of both the Total Solar Irradiance
(TSI, the irradiance integrated over the whole solar spectrum)
and the Spectral Solar Irradiance (SSI, irradiance integrated over finite spectral bands) using direct measurements or estimates (through proxies) 
of varying population of magnetic features on the solar disk \citep{domingo2009,ermolli2013}. While it is relatively well understood how 
sunspots dominate on a daily (sunspot evolution) or monthly (solar rotation) time-scale, the contribution of faculae and network, which dominate at the eleven-year
solar cycle and longer temporal scales, is still uncertain. This is in spite of numerous investigations of their photometric properties 
\citep[e.g.][]{foukal1985,ortiz2002,ermolli2007,berger2007,ermolli2010,viticchie2010,narayan2010,yeo2013,utz2013}, and attempts to reconcile observed properties of faculae 
and network with those predicted by models 
\citep{foukal1985,unruh1999,ermolli2007, ermolli2010}. 
As a result, while TSI reconstructions of the last and current cycles agree up to the 96$\%$ level, SSI
reconstructions, especially in the UV, and reconstructions of the past cycles obtained by different techniques, present discrepancies 
which are too large to accurately assess the effects of solar irradiance variations on the Earth atmosphere \citep[e.g.][]{ermolli2013,kopp2016, ball2016}. 


Reconstructions, in most cases, only partially take into account the variety of physical conditions which, as also inferred from numerical
modeling \citep[e.g.][]{spruit1976,pizzo1993,steiner2005,criscuoli2009,criscuoli2014}, determine the radiative emission of magnetic features. 
In particular, results obtained from the analysis of high spatial-resolution observations \citep{solanki1992,ishikawa2007,kobel2012, romano2012,feng2013} showed that 
small-size magnetic elements located in quiet regions are characterized by a higher photometric
contrast in photospheric continua, than magnetic elements located in active regions. 
\citet{criscuoli2013} employed three-dimensional magneto hydrodynamic (3D-MHD) simulations of the solar photosphere to show that such 
differences are caused by suppression of convection in active regions, which induces a decrease of the plasma temperature within and around magnetic elements.

\begin{table*}[t]
\footnotesize
\centering
\begin{tabular}{llcccc}
\hline\hline
Author(s)  & Brighter*  & Full disk &  Instr. (pixel size) & Criteria\\ 
\hline
\citet{ortiz2002} &  Faculae & Yes & MDI (1$^{\prime\prime}$)& B \\ 
\citet{ermolli2003} & Faculae & No & PSPT (2$^{\prime\prime}$), MDI (1$^{\prime\prime}$) & CaII, B \\
\citet{ishikawa2007}   &  Network &  No & SST/SOUP ($\leq$0.2$^{\prime\prime})$ & B\\ 
\citet{ermolli2007} & Faculae & Yes & PSPT (2$^{\prime\prime}$), MDI (1$^{\prime\prime}$) & CaII, B \\
\citet{kobel2011} &  Network & No &  Hinode/SP (0.3$^{\prime\prime}$) & B \\ 
\citet{romano2012} & Network & No & DST/IBIS ($\leq$0.17$^{\prime\prime}$)& AR prox.  \\
\citet{feng2013} & Network & No & DOT (0.07$^{\prime\prime}$) & G-band I$_c$  \\
\citet{yeo2013} &  Faculae & Yes & HMI (0.5$^{\prime\prime}$) & B  \\
$\dagger$\citet{yeo2014} &  Network & No &HMI (0.5$^{\prime\prime}$)& B \\
$\dagger$This paper & Network & Yes & HMI (0.5$^{\prime\prime}$) &AR prox. \\
\hline
Author(s)  & Brighter*  & Full disk & Source Code & \\ 
\hline
\citet{criscuoli2013}  &Network&No&Stagger (24 km)&- \\ 
\citet{riethmuller2017} & Network & No & MURaM (10 km) &-\\
\hline
\end{tabular}
\begin{tabular}{l}
* Brighter indicates that the feature has a higher intensity contrast for most field strengths and disk positions.\\
$\dagger$ These two papers represent studies in which the data are corrected for scattered-light. \\
\end{tabular}
\caption{\small A summary, albeit incomplete, of intensity contrast studies from both observations and simulations from the past fifteen years. 
The columns correspond to authors, year of publication, whether faculae or network were found to have a higher contrast at most disk positions and field strengths, 
type of observation, the instrument used and its spatial pixel scale and the criteria used to discriminate between network and faculae.
The bottom two rows correspond to simulation efforts. Therefore, no instrument is specified, instead a source code is named.  }
\label{table-sim-obs-compare}
\end{table*}

The high spatial-resolution results characterizing network as higher contrast features than faculae are apparently at odd with previous studies employing 
full-disk observations. For instance,  \citet{ortiz2002} analyzed data acquired with the Michelson Doppler Imager (MDI) aboard the Solar and Heliospheric
mission \citep{scherrer1995} and concluded that the contrast of network pixels is smaller 
and presents a rather modest center-to-limb variation with respect to the contrast of faculae. Similar results were found by \citet{yeo2013}, 
who analyzed data acquired with the Helioseismic and Magnetic Imager (HMI) aboard the Solar Dynamics Observatory \citep[SDO,][]{schou2012}, 
and by \citet{ermolli2003} and \citet{ermolli2007},  who analyzed photospheric solar images acquired with the Precision Solar Photometric 
Telescope \citep[PSPT,][]{coulter1994}. However, \citet{ortiz2002} and \citet{yeo2013} also noted that network elements are characterized by a higher 'intrinsic' 
contrast, thus suggesting that the lower contrast observed for network is mostly consequence of spatial resolution effects. 

Table 1 provides a summary of photometric contrast studies published in the last 15 years. This table is incomplete but gives the reader a  
comprehensive view of the findings of a variety of observational and numerical efforts \textit {at-a-glance}, including the lead author, 
year of publication, whether network or faculae were found to be brighter over the majority of field strengths and disk positions, whether or not the study was 
full-disk or limited in its field of view, the instrument used, its pixel size, and criteria for distinguishing between faculae and network.  

Different irradiance reconstruction techniques employ different approaches to take into account the contribution of magnetic features, but the majority of these 
techniques, included the two most employed in Earth-atmosphere studies, i.e. the Naval Research Laboratory models \citep[NRL][]{lean2000,coddington2016} 
and the Spectral and Total 
Irradiance REconstructions \citep[SATIRE][]{krivova2003,yeo2014b},  do not typically take into account the observational evidence 
of network elements being brighter than facular ones. 
In particular, NRL-TSI and NRL-SSI typically rely on measurements of the MgII and sunspot indeces 
to estimate total and spectral irradiance variations through multivariate analysis; in these models 
the contribution of network is therefore only indirectly accounted for through the derived correlation coefficients.
SATIRE models distinguish between the contribution of diferent magnetic structures, and the radiative emission of bright magnetic elements is assumed
to increase linearly with the magnetic flux, up to a saturation value \citep{krivova2003}, 
without taking into consideration the pixel's proximity to active regions. imilarly, other models not taking into account of a network 
brighter than faculae have been suggested in the literature \citep[e.g.][to name a few]{crouch2008,shapiro2011,morrill2011,thuillier2012,chapman2013, yeo2017}.
In particular, in SRPM 
\citep[e.g.][]{fontenla2011,fontenla2015}, OAR \citep[e.g.][]{ermolli2011} and COSI \citep[e.g.][]{haberreiter2008}, various types of magnetic features 
(which include different types of 
network and facular features) 
are classified according to their emission in chromospheric images (typically CaIIK), but the
modeled network contrast is lower than the facular one, especially toward the limb \citep[e.g.][]{ermolli2010}. Simplifications in implementing an irradiance reconstruction that 
accounts for all network and facular properties arise because reported 
differences are dependent on the methods of feature-discrimination used, and the intensity contrasts are a function of wavelength, magnetic field strength, 
spatial resolution of the instrument, activity levels and center-to-limb position \citep{solanki1993}.

We were inspired by \citet{yeo2014} who showed that the contrast-magnetic flux relation derived at disk-center using 
data compensated for the instrumental Point Spread Function (PSF) presents characteristics so far observed only using sub-arcsecond spatial 
resolution observations and simulations. We extend the \citet{yeo2014} work to include full-disk analysis of HMI data compensated for the PSF for
dates that sample a variety of magnetic activity on the disk. 

This paper adds to the existing literature on photometric contrasts of network and faculae as pertains to irradiance modeling in the following way:  
\begin{itemize}
\item {This is the first full-disk analysis to report on photometric contrast compensated for scattered-light of faculae and network defined by their proximity to 
active regions instead of their magnetic flux only.}
\item {We utilize HMI data corrected for scattered-light using a fast deconvolution routine already in the HMI pipeline 
\footnote{HMI data compensated for the PSF can be found within the HMI JSOC environment by searching for data series appended by '$\_$dcon'.
For example, HMI continuum intensity data obtained at a 45s cadence, normally designated as hmi. Ic$\_$45s, that have been corrected,
are found in the data series named hmi.Ic$\_$45s\_dcon. Several time periods of data are already available.
The HMI team is working towards
supplying data on a daily and continuous basis but the efforts are dependent upon funding outcomes.
Requests for scattered-light corrections for specific data periods and observables are welcome and should be addressed to $aanorton@stanford.edu$.}. 
The PSF was developed to account for both short- and long-range scattering range (see Sec.~\ref{sec:obs}).
The deconvolution is fast (less than 1 s per full-disk image) and could easily supply daily data for irradiance reconstruction purposes. 
Within this paper, we analyze original and PSF corrected full-disk intensity and magnetograms from ten different days between 2013 and 2015.}
\item {By utilizing a pre-existing HMI data-product, i.e. the HMI Active Region Patch data, we can distinguish between faculae and 
network in one step (see Sec.~\ref{sec:obs}). This simple methodology could easily be incorporated into ongoing irradiance reconstruction efforts. }
\end{itemize}

The analysis that we performed is similar to that of  \citet{ortiz2002} and \citet{yeo2013}, 
who analyzed data acquired with MDI and HMI, respectively. These previous studies did not differentiate between pixels located near AR, except that \citet{yeo2013}
excluded some pixels very close to active regions using a magnetic extension analysis, although after that exclusion there was no further distinction between,
or separate analysis of, pixels closer or further from AR. It is even noted in the conclusions of \citet{yeo2013} that  'while the largest effect is produced by 
the removal of magnetic signal adjoining to sunspot and pores...there remains a fair representation of active region faculae in the measured constrasts'.
Instead, network and facular regions were distinguished using the assumption that network and facular pixels are characterized by low/high
magnetic flux values, respectively. Moreover, the data employed by \citet{ortiz2002} were characterized by a different spatial resolution 
(about four times worse) and in both studies the data were affected by scattered-light. Therefore, we were inspired by \citet{ortiz2002} and \citet{yeo2013} 
to conduct a study on full-disk data while distinguishing faculae and network by their spatial proximity to AR.

\section{Observations and Data Analysis} \label{sec:obs}
We analyzed a set of 45 s Intensitygrams and Magnetograms acquired in ten different days between 2013 and 2015 with the  
HMI. The solar disk is imaged on a 4096$\times$4096 pixels detector, with a pixel/scale of 0.5 arcsec and a spatial resolution of 1 arc sec \citep{wachter2012}.
The HMI samples
the Fe I 6173.3~nm photospheric absorption line at six wavelength positions in two orthogonal circular polarization states.
The acquired filtergrams are then combined through an algorithm (the MDI-like algorithm) to produce estimates of the line-of-sight magnetic flux, Doppler velocity,
and Fe I  6173.34~{\AA} nearby continuum intensity, line-depth and line-width \citep{couvidat2012,cohen2015}. A full description of the HMI-pipeline is provided in 
\citet{couvidat2016}.
Estimates of the HMI observables are known to suffer from uncertainties resulting from various factors which include the assumption of a Gaussian shaped Fe I line profile,
saturation 
of the line in the presence of strong magnetic fields,
line-shifts induced by plasma motion, solar rotation and the orbital velocity of the spacecraft, 
stability of the tunable-filters and other optical components \cite[e.g.][]{fleck2011,liu2012,cohen2015,couvidat2016}.
The effects of these uncertainties on our results are discussed in Sec. \ref{sec:disc}.

Scattered-light is known to affect photometric studies \citep[e.g.][]{toner1997,mathew2007,criscuoli2008,yeo2014}. The PSF used for correcting HMI data for scattered-light is described in detail in Sec. 3.6 in \citet{couvidat2016}, 
so we limit our description to the basics. The form of the PSF is an Airy function convolved with a Lorentzian. The
parameters are bound by observational ground-based testing of the instrument conducted prior to launch \citep{wachter2012}, 
and by using post-launch, in flight data off the limb, during the transit of Venus and also during a partial lunar eclipse. The PSF employed herein 
is distinctly different from the PSF used by \citet{yeo2014}, which takes the form of a sum of Gaussian derived from the transit of Venus data. 
The PSF used by \citet{yeo2014} does not account for the large-angle, or long-distance scattering, since the shadow of Venus is too small to 
effectively measure the long-distance scattering. 
In addition, using the sum of simple functions such as Gaussian does not describe properly the diffraction-limited case, thus potentially introducing artifacts
on restored images \citep{wedemeyer2008}.  These can be avoided by introducing constraints on the parameters describing the width of the PSF \citep{yeo2014}.
For completeness, it must be mentioned that the PSF description as sum of Gaussian and Lorentzian functions is still proper for some applications, 
as for Earth-based observations dominated by seeing \citep[e.g.][]{toner1997,criscuoli2008}. 

To discriminate between pixels located in active and in network regions we employed the HMI Active Region Patches (HARPs) available for download from the 
Joint Science Operations Center (JSOC) website (\url{http://jsoc.stanford.edu/ajax/lookdata.html}), derived from HMI magnetograms following the procedure described in
\citet{turmon2002}. Note that the HARP mask locations do not change between the original and data compensated for scattered-light.
We discarded pixels belonging to sunspots umbrae and penumbrae by using an intensity
contrast criterion.
For each image we first estimated the average quiet-sun limb darkening as a function of the cosine of the heliocentric angle
$\mu$, by computing the intensity histograms at 60 different
$\mu$ values and averaging those pixels whose intensity values are within $\pm 3 \sigma$ the median value of the intensity distribution.
We created a contrast map to be the ratio of the intensitygram and the limb darkening image. We defined pixels as belonging to sunspots in places where the contrast is lower than four times the standard
deviation of the contrast image. Projection effects
were reduced by applying a 6-pixel kernel smooth on the obtained sunspot masks \citep{yeo2013}.

We also discarded from our analysis those pixels where the magnetic flux value is below three times the magnetogram noise level. This last quantity varied quadratically with $\mu$ and ranged from about 9 G at disk-center to about 12 G at the limb, as described in \citet{liu2012}.
Finally, due to the uncertainties toward the limb, we restricted the analysis to pixels located at $\mu>$~0.2.

\begin{figure*}
 \includegraphics[width=7cm,trim=0mm 0 50mm 0mm, clip=true]{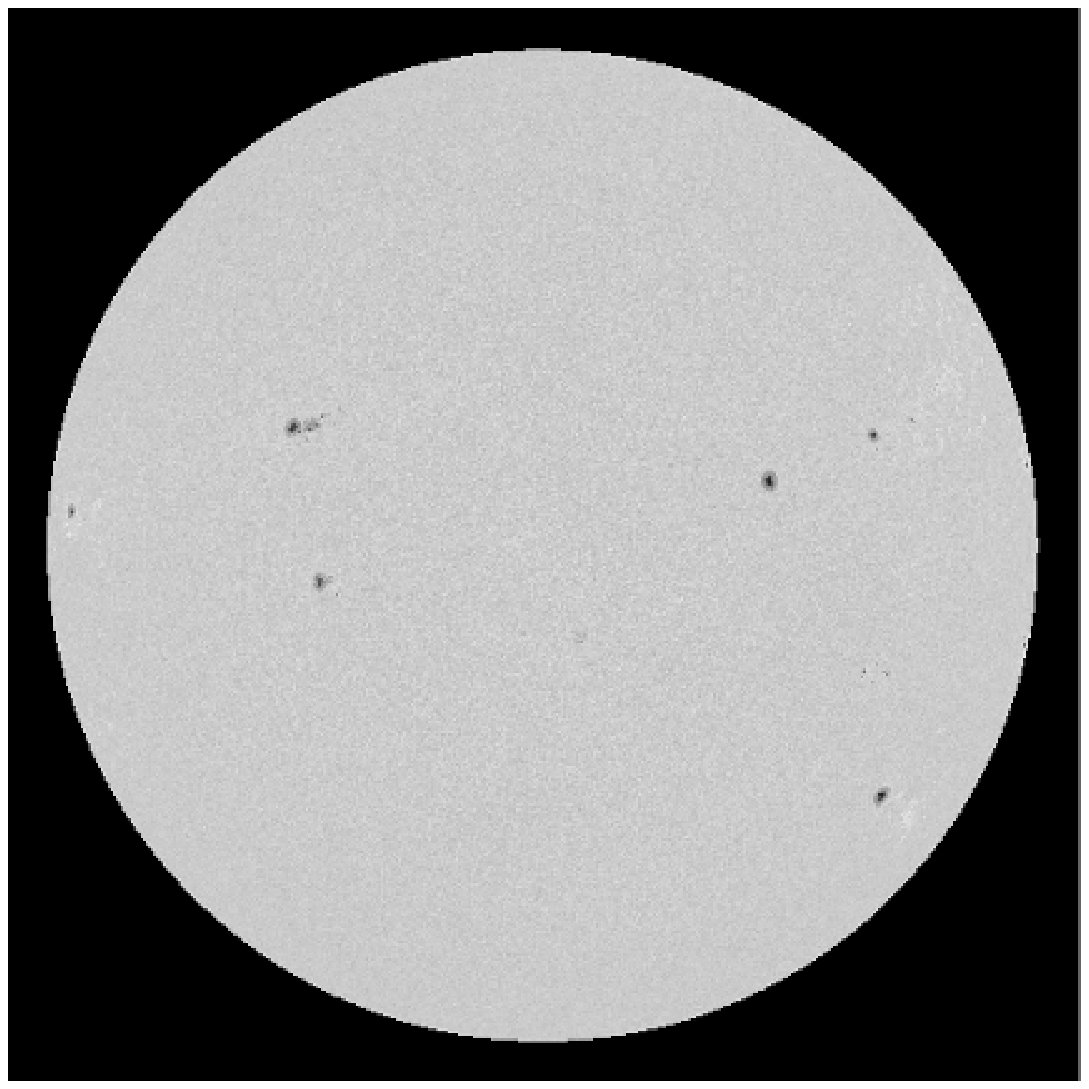}
 \includegraphics[width=7cm,trim=0mm 0 50mm 0mm, clip=true]{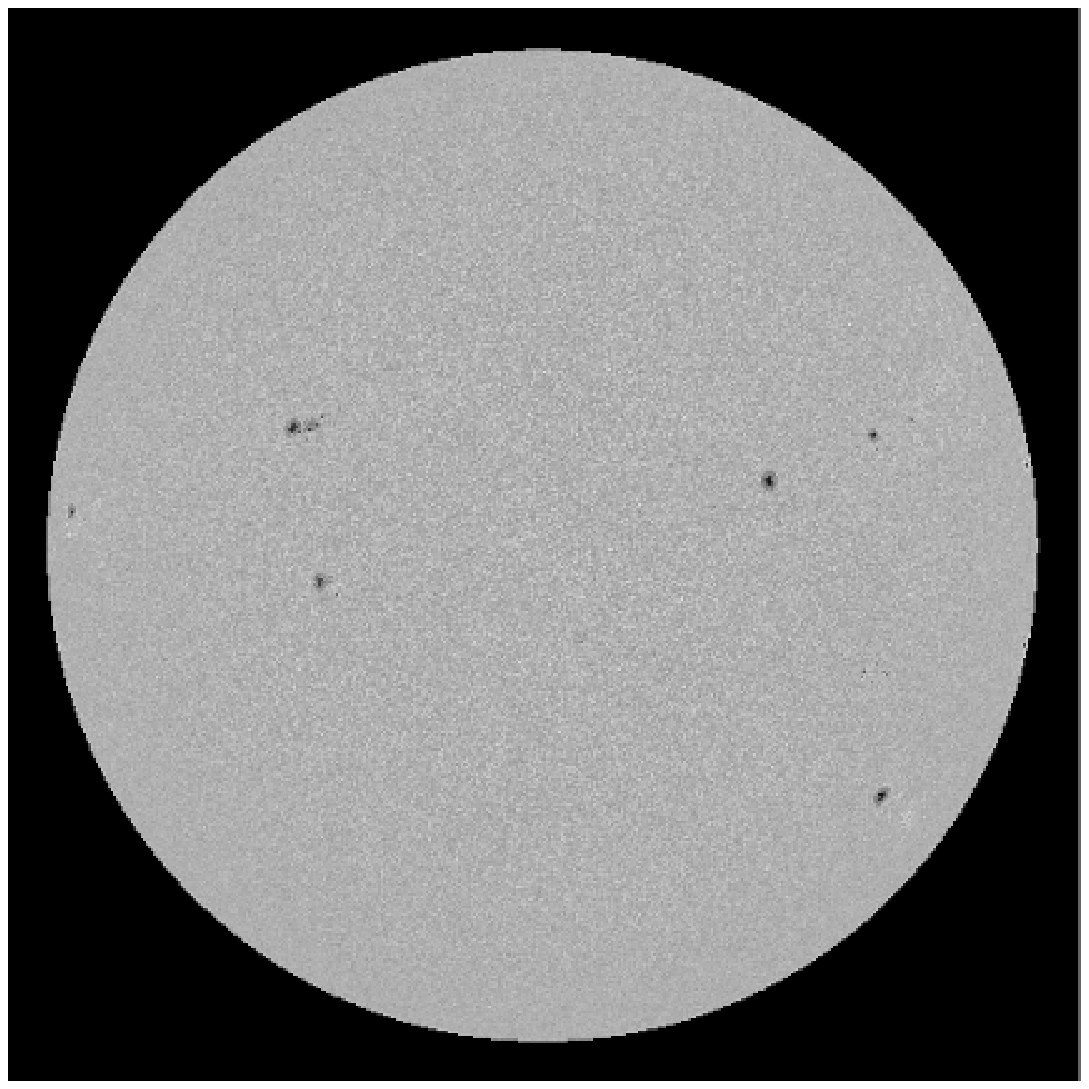}\\
  \includegraphics[width=7cm,trim=0mm 0 50mm 0mm, clip=true]{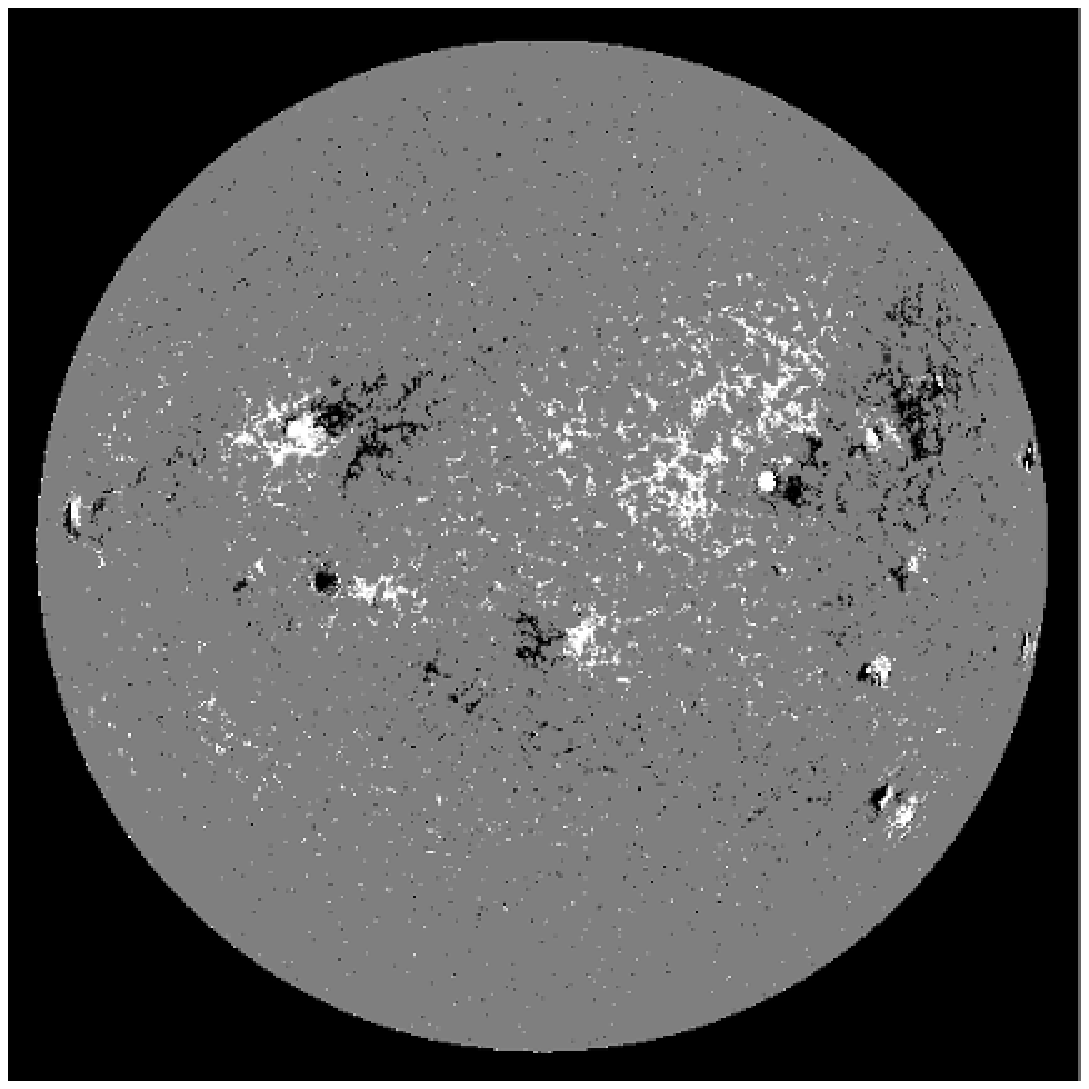}
 \includegraphics[width=7cm,trim=0mm 0 50mm 0mm, clip=true]{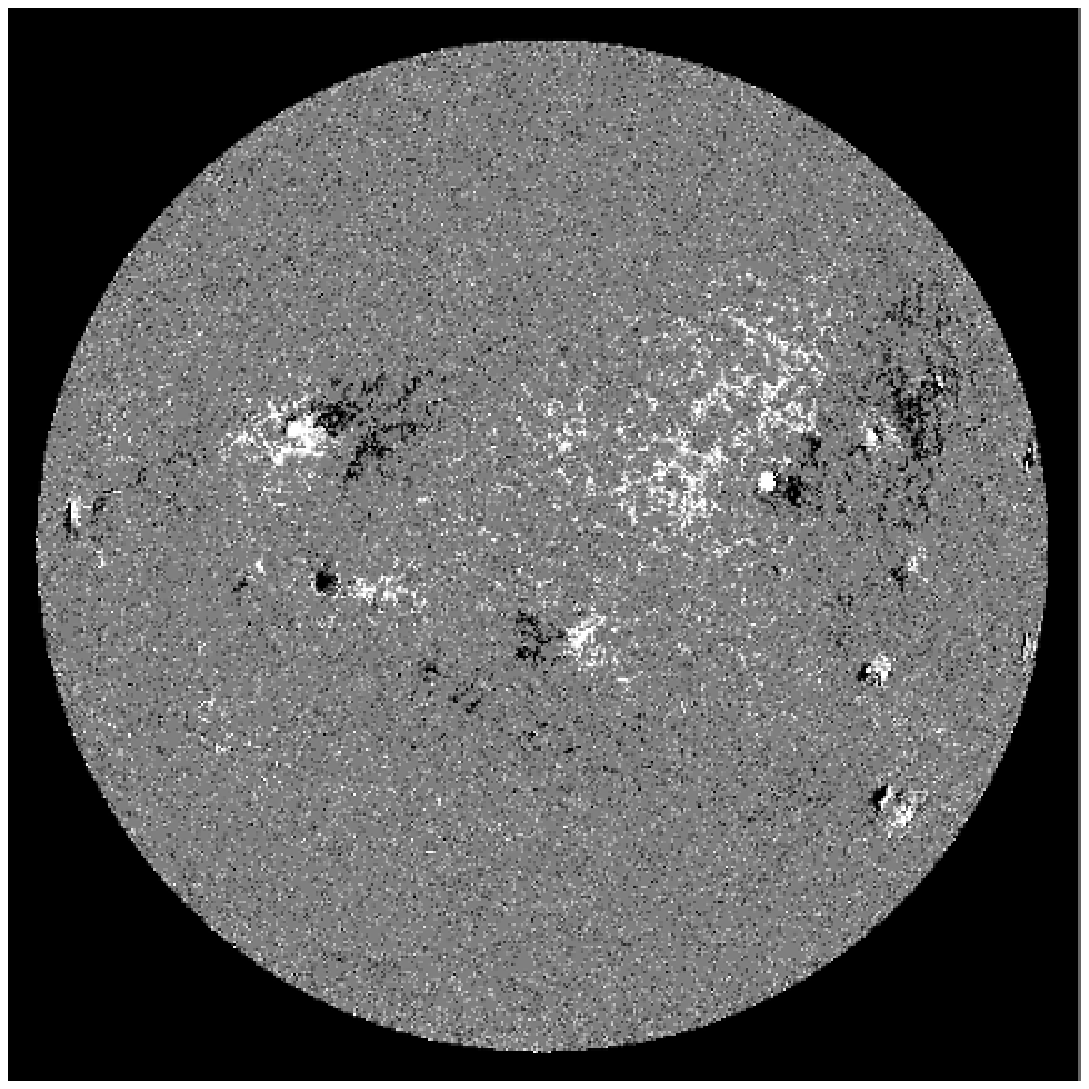}\\
  \includegraphics[width=7cm,trim=0mm 0 50mm 0mm, clip=true]{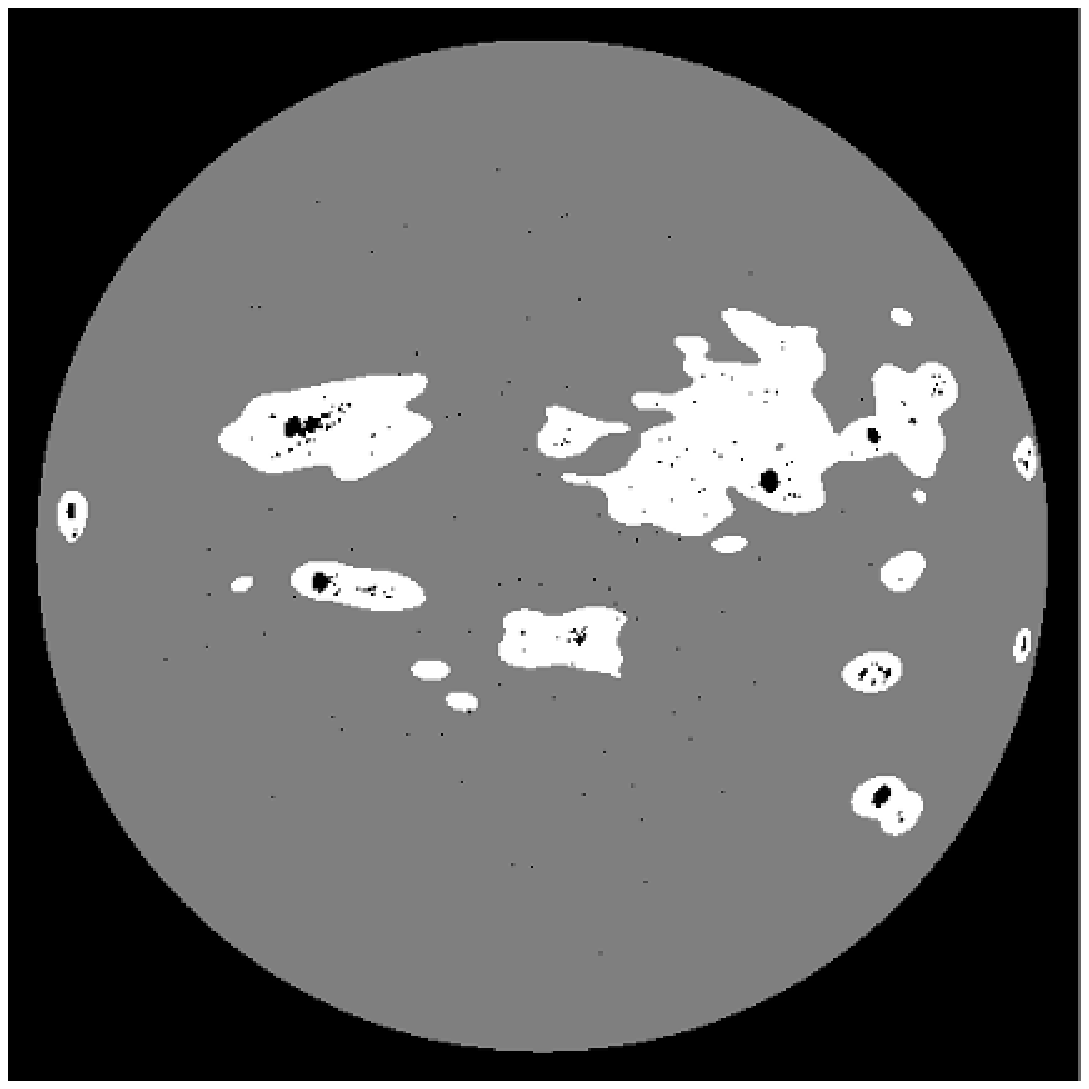} 
 \includegraphics[width=7cm,trim=0mm 0 50mm 0mm, clip=true]{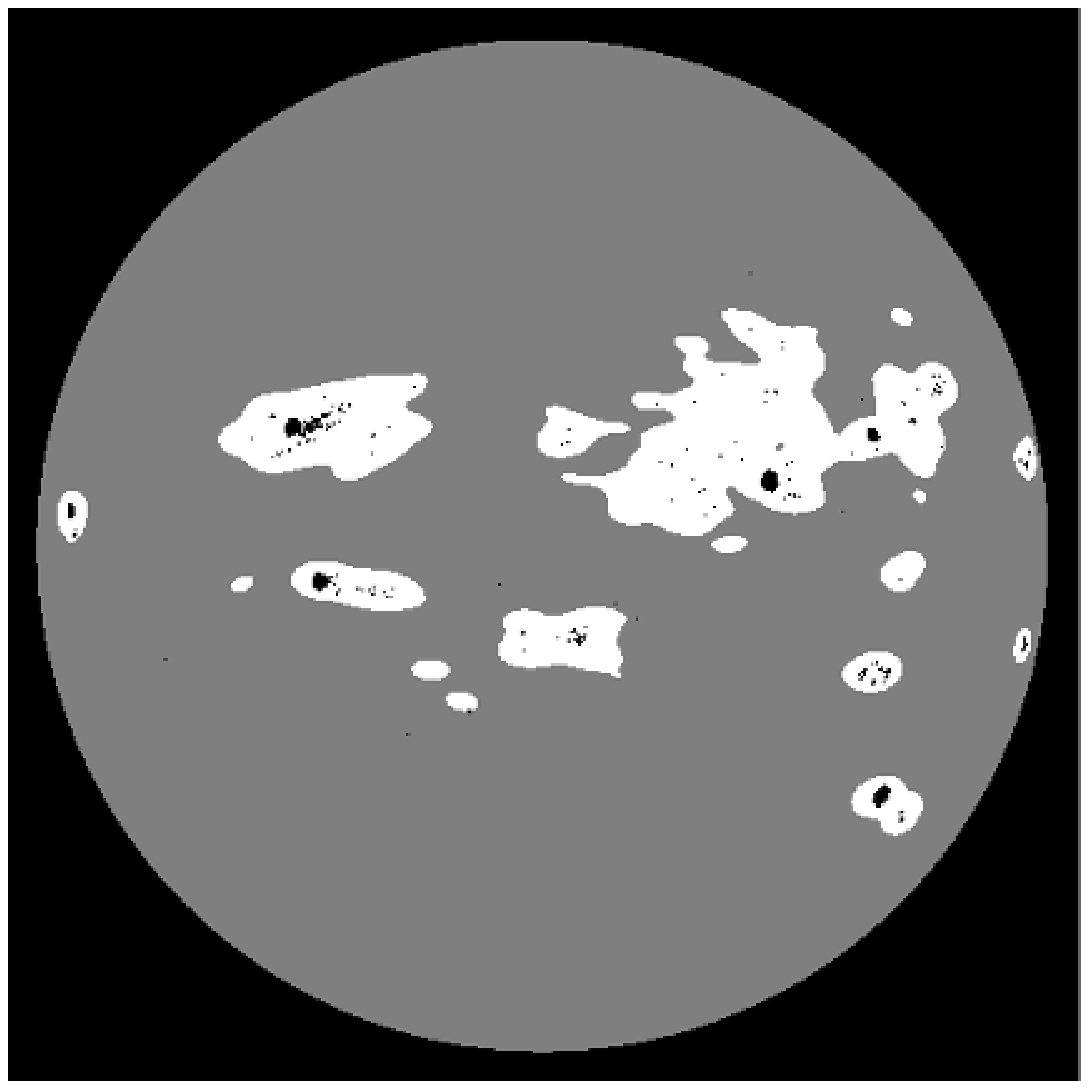}\\
 
 \caption{Examples of the analyzed data acquired on December 13th 2014. Intensitygram compensated for the limb darkening function (top), magnetogram
 saturated at
 $\pm$ 100~G (center) and masks (bottom) obtained on original (left) 
 and restored (right) data. The grey pixels on the magnetograms have line-of-sight magnetic flux below the noise threshold.
 The mask images show the pixels belonging to faculae (derived from HARP masks, see text) in white color, while black pixels on the disk belong to regions identified 
 as sunspot 
 umbrae, penumbrae and pores. 
 \label{fig1}}
\end{figure*}

Examples of a contrast map, a magnetogram (showing only pixels exceeding the noise level), the HARPs 
and sunspot masks are shown in Fig.~\ref{fig1}
for the original and the restored data. The images in Fig.~\ref{fig1} (bottom panels) show that the method applied to select sunspot regions also includes pores and micropores, and a small fraction
of dark intergranular lanes. These last type of pixels typically have magnetic flux lower than the magnetogram noise level and would have been discarded 
by the analysis anyways. Note that the difference in identifying dark features, as shown in the bottom panels comparing the left and right columns, arises from the standard deviation being lower in the original data compared to the compensated data. 
Meaning, more pixels in the original data were discarded as being lower than 4 sigma level.

We refer to pixels located within HARP regions as ``facular pixels'', while magnetic pixels located everywhere else on the disk will be referred to
as ``network pixels''. We use this notation because the first category of regions is more likely to include facular regions, whereas the second mostly 
include intranetwork and 
quiet and active network regions, but also because the method is straightforward to implement both in our efforts as well as in future irradiance reconstructions. 

\begin{figure*}
\begin{centering}
\includegraphics[width=7.7cm,trim=30mm 15mm 10mm 10mm, clip=true]{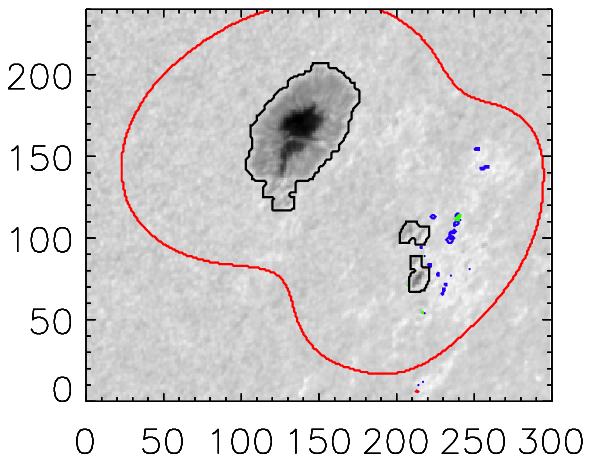}
 \includegraphics[width=7.7cm,trim=30mm 15mm 10mm 10mm, clip=true]{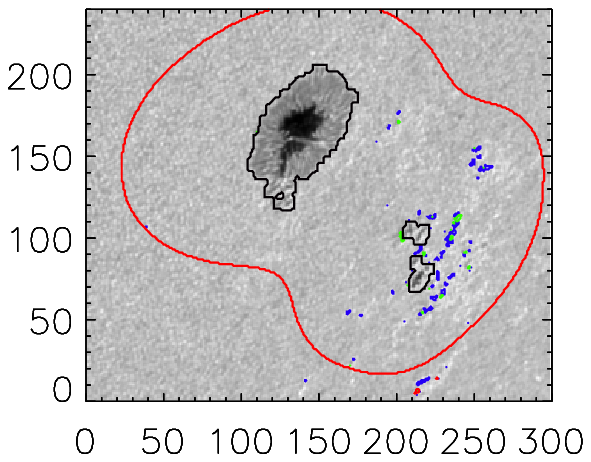}\\
\includegraphics[width=7.9cm,height=8.1cm,trim=25mm 15mm 10mm 10mm, clip=true]{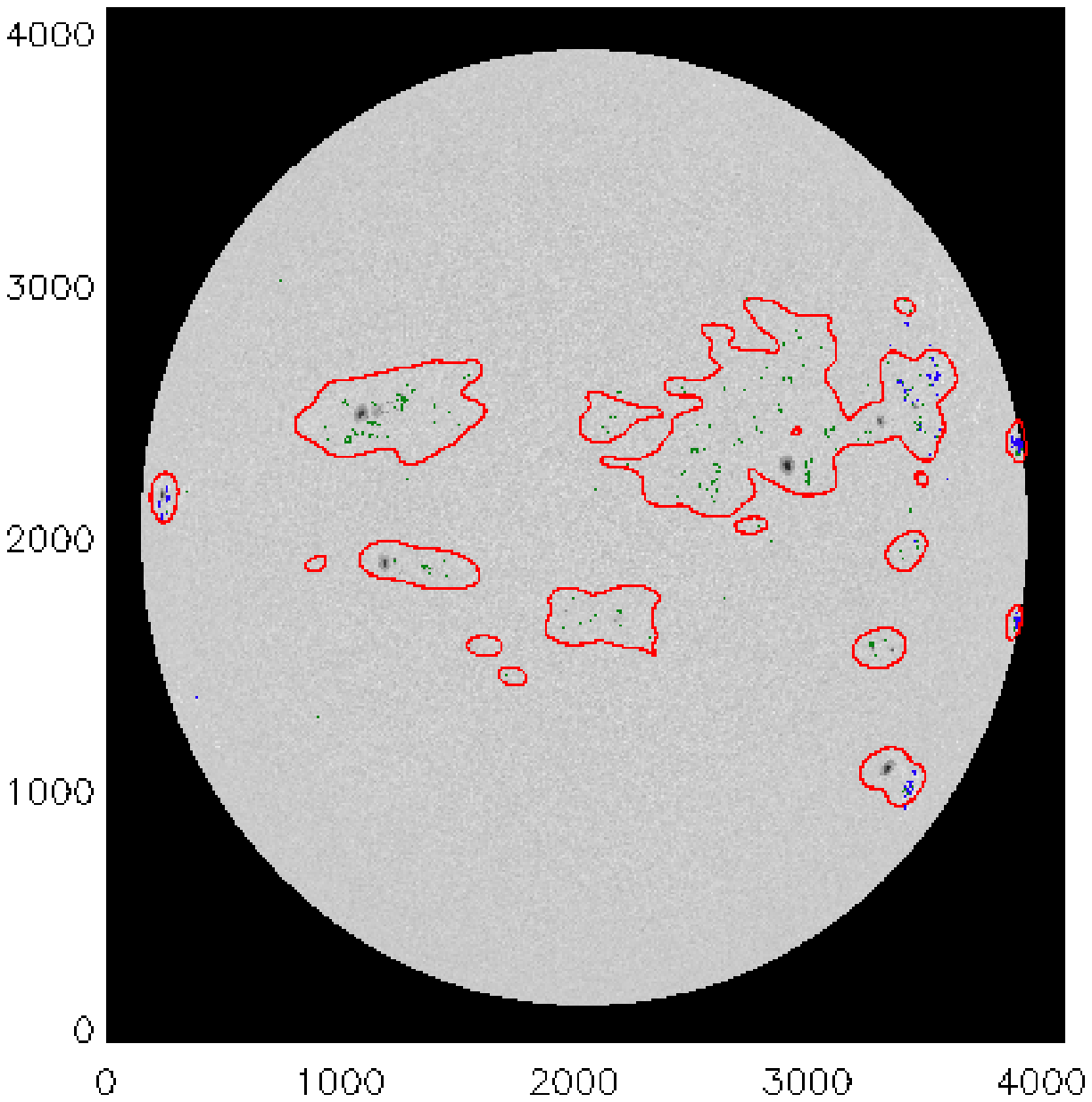}
\includegraphics[width=7.9cm,height=8.1cm, trim=25mm 15mm 10mm 10mm, clip=true]{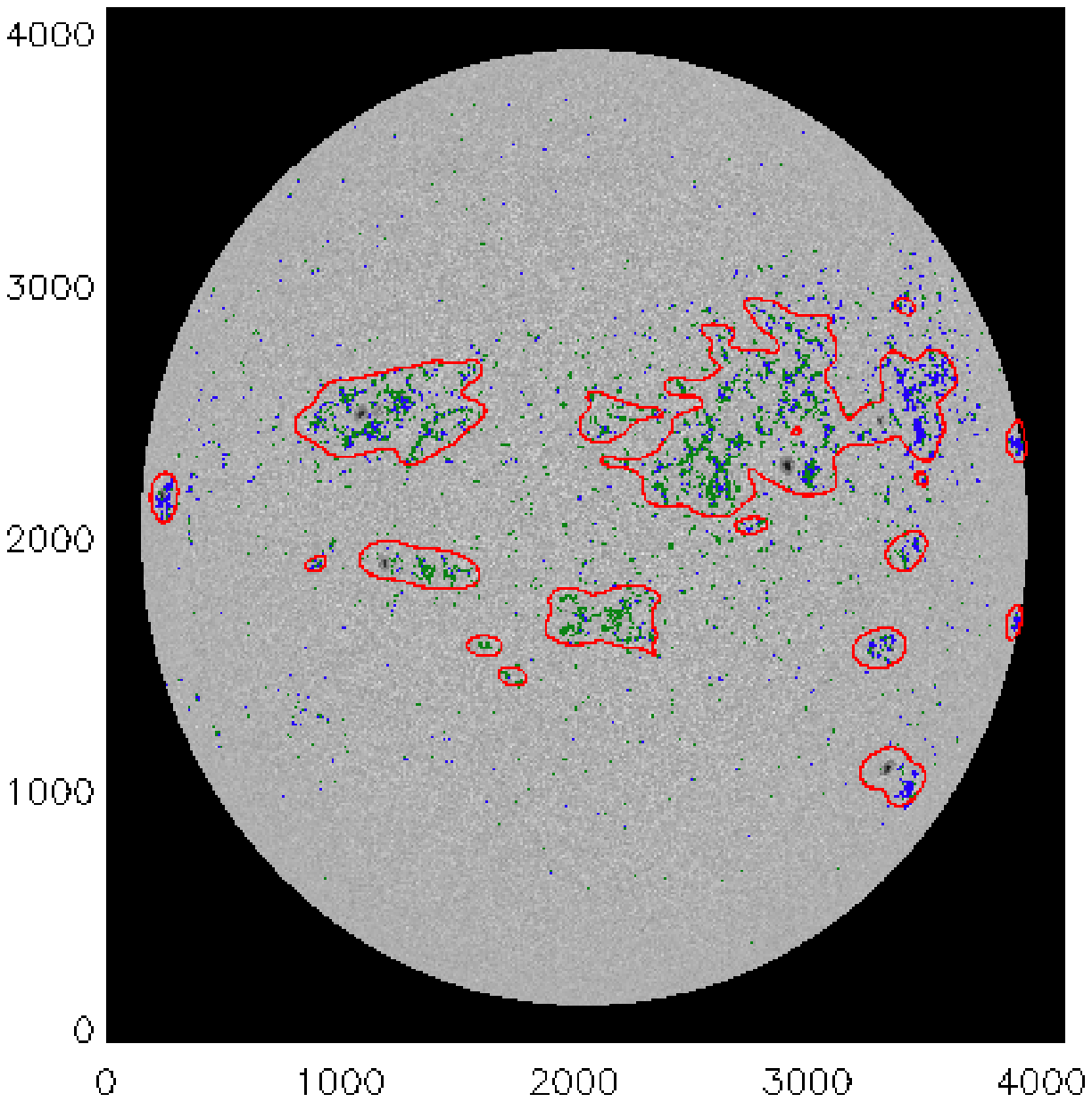}\\

 \caption{Top: Detail of the contrast image shown in Fig.\ref{fig1} for the original (left) and restored (right) data. Images show detail of AR12236 which, 
 on the day of the observation, was located at $\mu \simeq~0.4$.  Bottom: original (left) and restored (right) contrast images. The black contours in the top images
 enclose regions belonging to sunspots and pores, excluded by our analyses. In all images, red contours enclose the HARP regions, while blue and green lines enclose 
 pixels where $|B|/\mu >$~800~G and that appear brighter and darker, respectively, with respect to the quiet background.
 \label{fig2}}
 \end{centering}
\end{figure*}

It is important to note that \citet{ortiz2002} and \citet{yeo2013} considered only pixels where the magnetic flux compensated
for line-of-sight effects ($|B|/\mu$) is lower than 800~G, the number of large magnetic flux pixels being statistically not significant in their data. 
We decided instead to extend the analysis to higher magnetic flux values, as the restoration increases the number 
of large magnetic flux pixels (see Sec.~\ref{sec:solairra}).
\citet{yeo2013} also argued that high magnetic flux values, especially toward the limb,  most likely result from horizontal fields, and that these are typically 
associated with sunspots and pores. Inspection of our data confirms that statistically this is the case for pixels located at the extreme limb ($\mu<0.1$), which are
discarded from our analysis, but we do not find a clear association of 'dark' pixels 
within locations of sunspots and big pores at other positions over the disk. 
Figure ~\ref{fig2} shows for instance, in blue and green color, pixels with magnetic flux
larger than 800~G, which appear brighter and darker, respectively, with respect to the average quiet sun intensity, belonging to a HARP Region located at $\mu\simeq~0.4$. 
On the contrary, the bottom panel of Fig.~\ref{fig2} shows that, especially on restored images, these pixels seem to be distributed everywhere 
over the disk, 
with a preference in active regions and in the activity belt, but not exclusively at disk-center. 

\section{Results}\label{sec:res}
For each pixel $i$ over the disk we defined the continuum intensity contrast as $C=\frac{I_i}{I_q} -1$, where $I_i$ is the continuum intensity of the pixel, and $I_q$
is the average quiet sun intensity of pixels located at similar angular distance from disk-center, estimated as described in Sec.~\ref{sec:obs}.
We investigated the dependence of the contrast 
on the magnetic flux, B, and on the cosine of the heliocentric angle, $\mu$.
We considered 50 G bins of magnetic flux values compensated for line-of-sight effects ($|B|/\mu$) of pixels located at
16 different radial distances from disk-center, in $\Delta \mu=0.025$ intervals.

\subsection{Comparison between results obtained on original and restored data}\label{sec:deconv}
In this section we discuss the effects of restoration on the determination of the dependence of the intensity contrast on the magnetic flux and on the cosine of the 
heliocentric angle. We refer the reader to \citealt{yeo2014} for an additional description of the effects of the compensation for the instrumental 
Point Spread Function on the HMI observables.

\begin{figure*}[h]
\includegraphics[width=16cm,height=10.6cm, trim=3mm 4 1mm 4mm, clip=true]{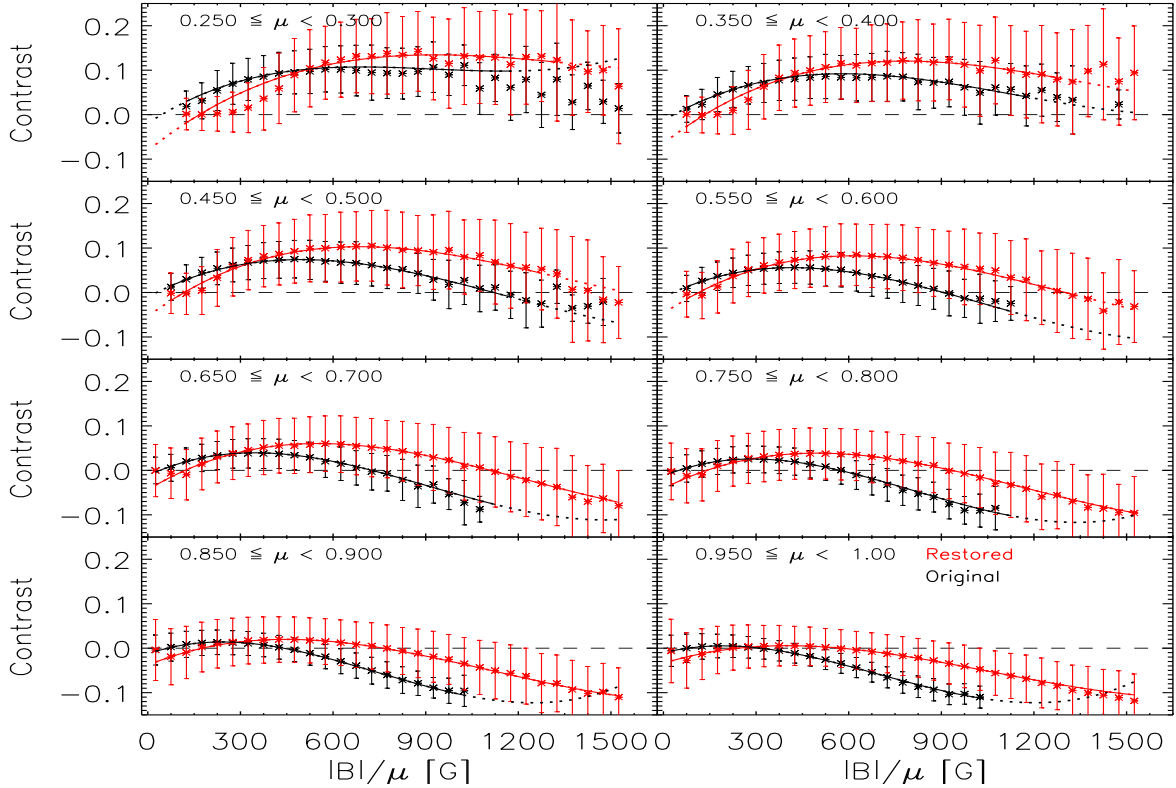}
 \caption{Variation of the intensity contrast with the magnetic flux for pixels located at various radial distances from disk-center derived by original (black)
 and restored data (red). Continuous lines: cross-sections of the surface fits to the data (see Sec.~\ref{fits}); dotted lines: 
extrapolation of the fit.
 \label{fig3}}
\end{figure*}

\begin{figure*}[h]
\includegraphics[width=16cm,height=10.6cm, trim=3mm 4 1mm 4mm, clip=true]{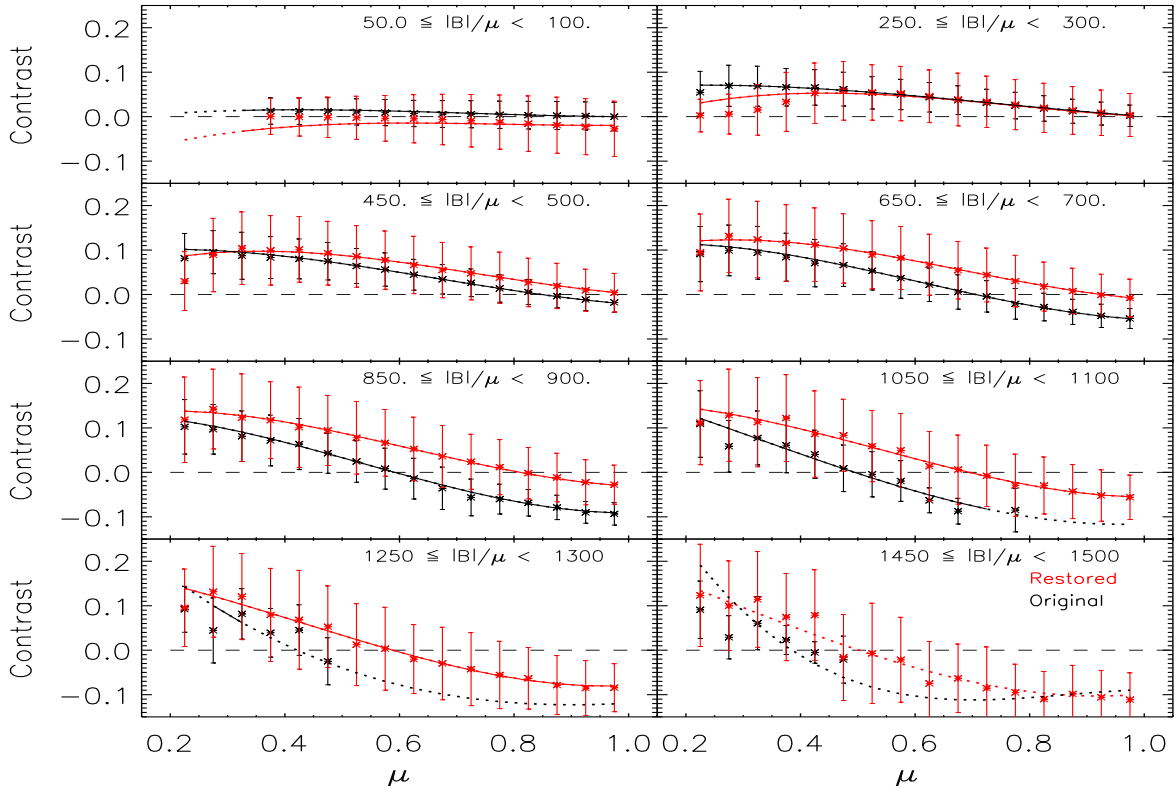}
 \caption{Variation of the intensity contrast with the cosine of the heliocentric angle for various magnetic flux ranges derived by original (black) 
 and restored (red) data. Continuous lines: cross-sections of the surface fits to the data (see Sec.~\ref{fits}); dotted lines: 
extrapolation of the fit.
 \label{fig4}}
\end{figure*}

Fig.~\ref{fig3} shows how the intensity contrast depends on the magnetic flux value at eight different heliocentric angles for original and restored data. Fig.~\ref{fig4} shows, instead, the variation of the intensity contrast with $\mu$ for eight magnetic flux ranges.
In both plots, data points correspond to average intensity contrast values computed over the corresponding bins 
while the error bars represent the standard deviations of values in each bin.

Fig.~\ref{fig3} shows that the contrast, towards disk-center, is negative for magnetic flux values smaller than about 200~G, but the contrast increases to reach 
a maximum between 300-400~G and then decreases again. This ``fish hook'' trend has been observed  
in the analysis of sub-arc second spatial 
resolution observations \citep[e.g.][]{schnerr2011,kobel2011,kobel2012,kahil2017}, and was also obtained by \citet{yeo2014} on HMI data compensated 
for the instrumental PSF
(a detailed comparison with the results obtained by these authors is given in Sec.~\ref{sec:disc}). 
The ``fish hook'' trend is not seen in the original data, and is partially visible only when increasing the magnetic flux bin size (see Sec.~\ref{sec:disc}).

\citet{rohrbein2011} employed 3D MHD simulations to interpret the physical origin of the contrast-magnetic field dependence observed at disk-center. They showed that
the decrease of contrast for low magnetic flux values is consequence of the accumulation of the flux within intergranular lanes, while the decrease of contrast
at large magnetic flux values is an effect of reduced spatial resolution in observations, which decreases both the magnetic flux and the contrast of the bright edges of 
micropore structures. This also explains why the restoration (Fig.~\ref{fig3} and Fig.~\ref{fig4}), contrary to what typically expected,
toward disk-center produces a (small) enhancement of the average contrast only at 
the lower magnetic flux values, while at higher magnetic fluxes the contrast of restored pixels is up to 4-5 times smaller. 

Toward the limb the restoration increases the contrast for $|B|/\mu >$ 300 G, 
as expected by the fact that the contrast at small values of $\mu$ is determined by the ''hot wall`` effect. The amount of contrast variation induced 
by the restoration is
function of both the angular distance and the magnetic flux, generally increasing with the magnetic flux and toward the limb, where the contrast of restored pixels is up 
to 4-5 times larger than the original ones. 
On the other hand, the restoration seems to have little effect 
on the angular distance at which the contrast reaches the maximum, while it shifts toward disk-center the value of $\mu$ at which magnetic elements become brighter 
than the background. This result is explained by the fact that the restoration compensates for scattered-light effects, as 
previous studies indicate that the position of the maximum contrast is sensitive to the spatial resolution \citep[e.g.][]{criscuoli2009},
while the angular position at which features become brighter is less so \citep[e.g.][]{yeo2013}. 

Finally, as noted in previous studies \citep[e.g.][]{criscuoli2008}, the standard deviation in each bin is enhanced by the restoration. In particular,
as also noted in Sec.~\ref{sec:intro} (Fig.~\ref{fig2}), although on average the restoration causes highest magnetic flux features to appear ``brighter'', there is still a considerable amount 
of features that instead appear ``darker''.

\begin{figure*}
\includegraphics[width=16cm,height=10.6cm, trim=3mm 4 1mm 4mm, clip=true]{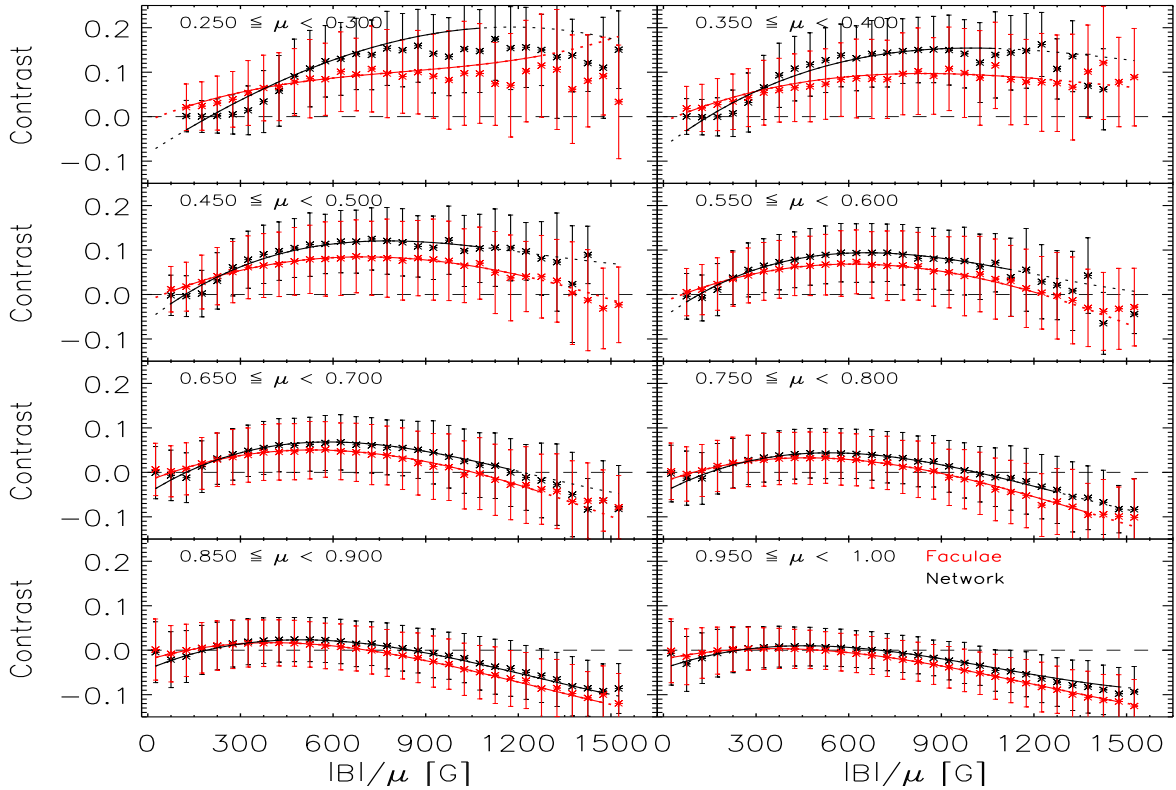}
 \caption{Variation of the intensity contrast with the magnetic flux for pixels located at various radial distances from disk-center in facular (red)
 and network (black) regions singled out on restored data. Continuous lines: cross-sections of the surface fits to the data (see Sec.~\ref{fits}); dotted lines: 
extrapolation of the fit.
 \label{fig5}} 
\end{figure*}

\begin{figure*}
\includegraphics[width=16cm,height=10.6cm, trim=3mm 4 1mm 4mm, clip=true]{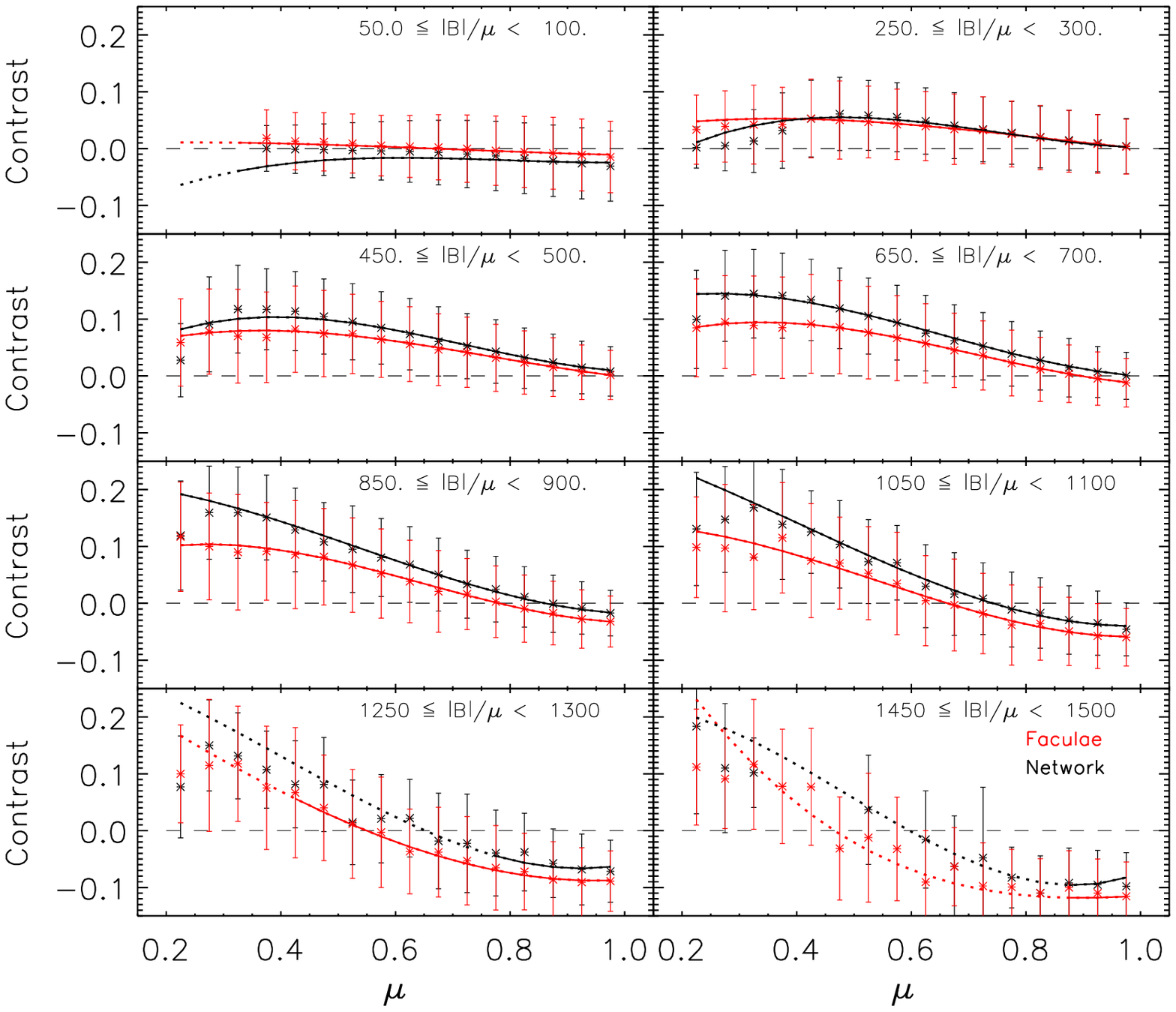}
 \caption{Variation of the intensity contrast with the cosine of the heliocentric angle in
 facular (red) and network (black) regions singled out on restored data. Continuous lines: cross-sections of the surface fits to the data (see Sec.~\ref{fits}); dotted lines: 
extrapolation of the fit.
 \label{fig6}}
\end{figure*}

\subsection{Comparison between Network and Faculae}
The differences between network and facular regions are illustrated in Fig.~\ref{fig5} and Fig.~\ref{fig6} for results obtained from 
 restored data only. Results obtained from original data show little or no difference of contrast between pixels located in different magnetic activity
regions, especially at disk-center, and are shown in Appendix~\ref{A1}. On the contrary, results obtained from restored data show that
pixels located in intergranular lanes, or at low magnetic flux values (about less than 300~G),  appear always darker in facular regions rather than in network ones.
Pixels with higher magnetic
flux are instead always brighter in network regions. We also note that the contrast differences are smaller toward the disk-center
but increase toward 
the limb, where the network contrast at higher magnetic fluxes can be up to almost twice the facular one. Finally, inspection of Fig.~\ref{fig6} reveals that the angular position of the maximum contrast is similar for pixels located 
in different regions, while
the angular positions at which the contrast changes sign occur closer to disk-center for network rather than for faculae.  

These results are in qualitative agreement with results obtained
by the analysis of sub-arc second spatial resolution observations at disk-center \citep{ishikawa2007,kobel2011, kobel2012, romano2012, feng2013}. 
As explained in \citet{criscuoli2013} the different contrasts obtained
in active and in quiet regions are consequence of the decrease of temperature in photospheric layers induced by the suppression of convective motions 
within active regions. This reduces the contrast of 
granulation (both granules and intergranular lanes) and consequently of small-size magnetic flux concentrations whose temperature stratification is determined by the 
temperature of the surrounding plasma. Similarly, toward the limb the observed higher contrast in network regions can be explained by the reduction of
the 'hot wall' temperature in facular regions. \\
Finally, a comparison of plots in Fig.~\ref{fig4} and \ref{fig5} with those in Fig.~\ref{fig1A} and \ref{fig2A} reveals that the effects of restoration are
larger for network pixels, where, for the largest magnetic fluxes, the contrast variations can be up to twice the one obtained for faculae.

\subsection{Surface fits} \label{fits}
\begin{figure}
\includegraphics[width=8cm, trim=3mm 4 1mm 4mm, clip=true]{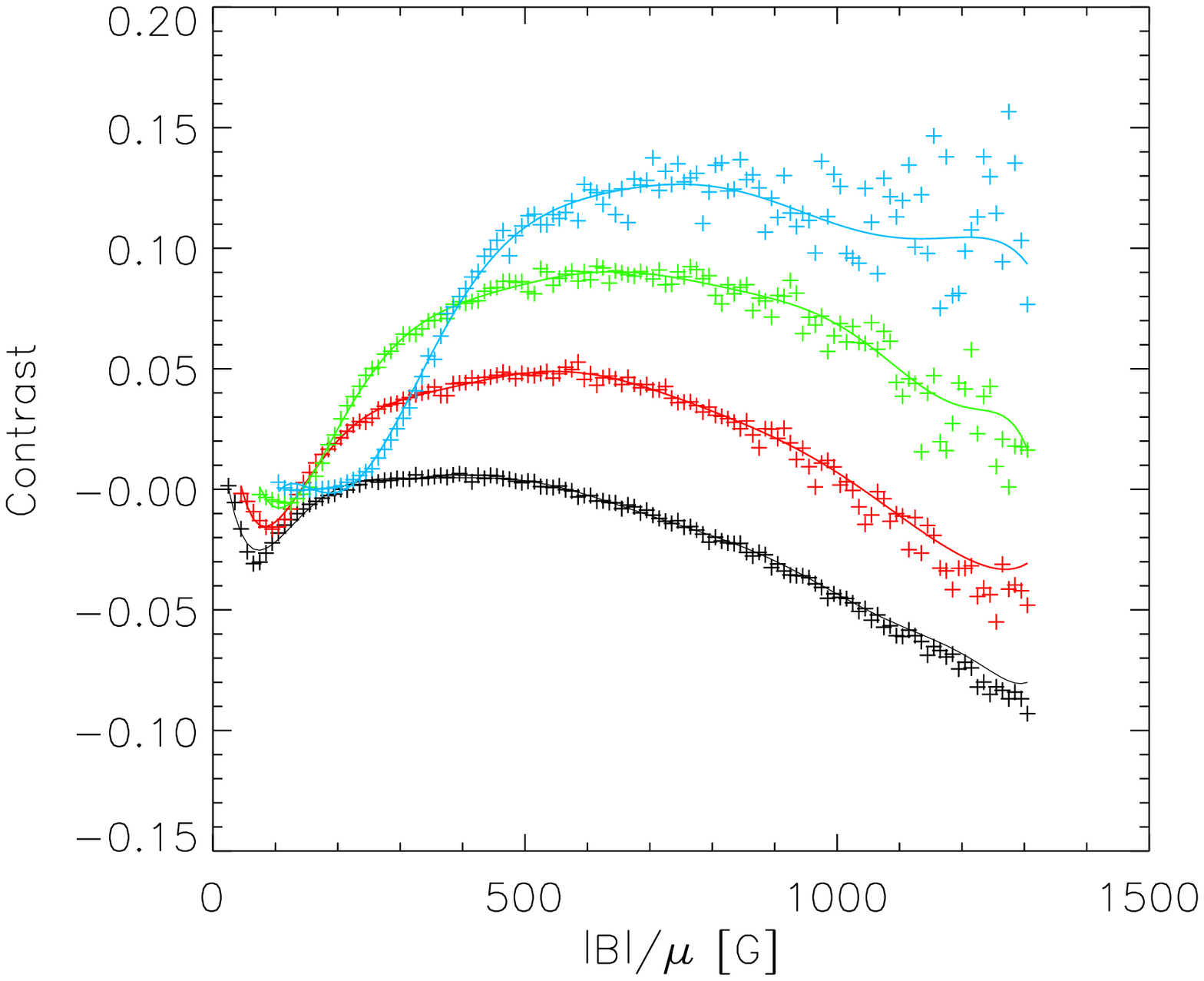}
 \caption{Comparison between results obtained from restored data (plus signs) and 10th-order polynomial fit to the data (continuous lines).  Black: $\mu$=0.975; red: $\mu$=0.725;
 green: $\mu$=0.525; blue: $\mu$=0.325.
 \label{fig6a}}
\end{figure}

Following \citet{yeo2013}, we fitted our results with cubic surface functions. The analytical form of the fit and the values 
of the fit coefficients 
 are reported in Appendix~\ref{A2}. The fits 
were evaluated using data binned over 16 equally spaced in $\mu$ values and 10~G magnetic flux intervals. 
Bins where the number of elements was lower than 100 were excluded 
from the analysis by imposing a null weight during the fit.
The results are illustrated in Figs.~\ref{fig3},
\ref{fig4}, \ref{fig5} and \ref{fig6}.
The agreement of the fits with the observational data points is in general very good at most of the flux values and angular positions, especially at the data
points employed to produce the fits.
The agreement is worse at small magnetic flux values on restored data, where the fits do not reproduce the
'fish hook' shape. Recently, \citet{schnerr2011} modeled the contrast-magnetic field relation obtained from the analysis of SST \citep{scharmer2003} 
and HINODE \citep{kosugi2007}
data as a sum of 
exponential functions,  while \citet{kahil2017}, who analyzed SUNRISE data \citep{barthol2011},  employed a logarithmic function, although in this last case the fit reproduced
the contrast at magnetic flux values larger than approximately 80~G (that is the dimming at small magnetic flux values was not reproduced). We found that 
the contrast-magnetic flux relation obtained from restored data is best reproduced when using a 10-th order polynomial in $|B|/\mu$.  Figure~\ref{fig6a} shows indeed
that this function 
reproduces the observations at both low and high magnetic flux values at various angular positions. Results from the fit are reported in the Appendix~\ref{A3}. 
 
\section{Discussion}\label{sec:disc}

\subsection{HMI uncertainties}\label{sec:disc:hmi}
Previous investigations showed that 
HMI data-products are affected by uncertainties stemming from the pipeline employed to estimate the data-products as well as by instrumental effects 
(see Sec~\ref{sec:intro}). \citet{cohen2015} employed results from numerical synthesis of the Fe I 617.3 nm line to show that the error in the estimate of the 
continuum intensity introduced by approximations in the MDI-like algorithm are function of the magnetic field and the line-of-sight, and, as second order effects, of 
Doppler shifts. Specifically, their Fig.~3 shows that the uncertainty for a facular region with associated a magnetic field of 1000~G is lower than 2\% at disk-center 
and decreases rapidly toward the limb, while the uncertainty for quiet regions increases  toward the limb up to about 1\%. Therefore  
uncertainties in the estimates of the contrast induced by approximations in the MDI-like algorithm are below 2\% and show little dependence with $\mu$. These uncertainties 
are typically below the amplitude of the error bars in our plots, and we therefore conclude that they have negligible effects on our results.

Because we used the contrast to characterize the photometric properties of magnetic elements, the increase of opacity of the entrance window during the first years 
of operations of the SDO, also reported in \citet{cohen2015}, have no influence on our results. Uncertainties stemming from variations of 
the properties of the filter transmission profiles are instead more 
difficult to assess, as there is no direct measurement available. \citet{criscuoli2011} estimated uncertainties lower than 10\% introduced
by variations of the transmission profiles
for continuum measurements provided by the MDI, which, 
similarly to HMI, 
combines filtergrams acquired with a tunable filter. In the case of HMI we expect uncertainties due to these effects to be smaller, as the
filters are periodically retuned. Moreover, the data analyzed were acquired in a relatively short temporal frame (about three years), so that
effects introduced by variations of the positions and shapes of the filters are most likely negligible on our results. 

In addition, \citet{cohen2015} showed that HMI estimates of the Fe I 617.3 nm line core intensity are affected by large uncertainties (several tens of percent). 
For this reason we refrained from using those HMI data-products, although such measurements would have provided extremely valuable information about the different 
temperature stratification within network and facular structures.

It is also important to mention that similar studies carried out on magnetograms usually employed data averaged over longer temporal range 
\citep[][employed for instance HMI data averaged over 312 s.]{yeo2013} to reduce noise and p-mode oscillation effects. In our analysis we decided to employ 45 s data.
To investigate how much of a difference it might make to use 720 s data instead of 45 s data, we repeated the analysis on a subset of three 
720 s images and found that the curves describing the dependence of the contrast on the magnetic flux agree within the 2\% level (and are therefore not reported).
Hence, we conclude that our results are not affected by the use of different types of HMI data. 

Finally, the Lucy-Richardson algorithm is known to potentially enhance noise in restored data \citep{white1994}. Inspection of the restored data (e.g. Fig.~\ref{fig2}) suggests this effect to be negligible. An analysis of the power spectra of both intensity images and magnetograms reveals indeed a small enhancement
of power beyond the frequency cutoff of the telescope, but the cumulative power beyond this frequency is below 2\%.
To investigate the effect of this enhancement on our results we first estimated the noise 
level on a subset of original and restored magnetograms employing a method similar to the one described in \citet{liu2012} (we removed active pixels from the analysis while
\citealt{liu2012} analyzed data acquired during low magnetic activity), and found that the restoration typically doubles the noise level. We therefore repeated the 
analysis increasing the noise level threshold by a factor of two and found  that the maximum relative difference between results obtained with the two noise thresholds 
is below 1.8\%.

\subsection{Features classification and quiet Sun definition}
It is very well known that identification methods employed to singled out features on solar images can affect the derived properties of such features \citep[e.g.][]{ermolli2007,jones2008,yeo2013,ulrich2010,ashmari2015}.
We employ masks produced with the method suggested in \citet{turmon2002} to distinguish between facular and network regions.
As explained in Sec.~\ref{sec:intro}, the HARP regions employed  mostly
include faculae and probably part of features that other methods might have classified as active network. Magnetic pixels not-belonging to HARPs
include intranetwork, network and active network and no effort was made to discriminate between these latter types of features.
On the other hand, the adopted classification is sufficient for the purpose of this study, 
which is to investigate the effects of the level of activity of the surrounding plasma on the photometric contrast of magnetic features. 

It is important to notice that in order to further reduce the effects of noise, some authors apply a minimum size threshold (typical between 1 to 10 pixels) 
to the structures analyzed
\citep[e.g.][]{ortiz2002,jin2011,yeo2013,criscuoli2016}. To investigate the effect of isolated pixels on the estimated average contrasts, we then repeated the analysis 
applying an ``opening'' operation with a 2-pixels kernel to the pixels exceeding the noise level of the magnetograms \citep[see e.g.][]{criscuoli2016}. We found that 
the application of such threshold produces a modest increase of the average contrast of pixels with low magnetic flux located toward the limb, with the largest effects found 
on restored data. 
In particular, for pixels with $|B|/\mu \le 300~G$ and $\mu \le 0.5$ singled out on original data the average and maximum relative increase of contrast are  0.6\% and 2.6\%, 
respectively, while for pixels singled-out on restored data are 1.7\% and 3.2\%, respectively. These differences are well below the statistical uncertainties of our measurements,
so that we can conclude that the application of a minimum size threshold on our data does not affect our results.

Photometric contrast is also known to be affected by the arbitrary definition of quiet sun regions \citep[e.g.][]{peck2015}. 
We therefore compared our results with those obtained defining as quiet pixels those that are below the noise level on magnetograms. 
The contrast relative differences found with this method and the one described in Sec.~\ref{sec:obs} is below 0.6\%, so that we conclude that our results
are not significantly affected by the criteria adopted to define quiet sun pixels.

\subsection{Comparison with previous studies}\label{sec:disk:sec:comp}
The investigation of the dependence of magnetic elements' continuum contrast with the heliocentric angle and the magnetic flux has been the subject of several studies 
(see Sec.~\ref{sec:intro}). Qualitatively in agreement with previous analysis \citep{foukal1990,topka1997,sanchezcuberes2002,ortiz2002, walton2003,ortiz2006,yeo2013}, 
 we found that highest magnetic flux features appear dark at disk-center and 
bright toward the limb, the value of  $\mu$ at which the contrast changes sign depending on the value of the magnetic flux considered.
However, a quantitative comparison of our results with these studies is hampered by the different observing conditions, as spatial resolution, scattered-light,
wavelength
and methods employed to identify the magnetic features 
play an important role in determining the dependence of contrast on magnetic flux and line-of-sight \citep[e.g.][for a review]{criscuoli2009}.

\begin{figure}
\includegraphics[width=8cm,height=6cm,trim=1mm 0 2mm 0mm, clip=true]{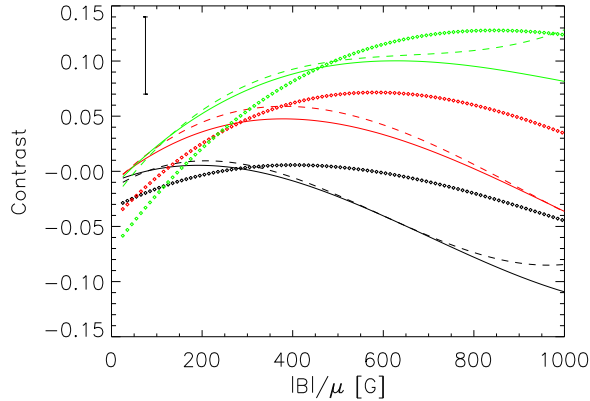}
 \caption{Comparison of contrasts obtained from cubic surface fits on original data (continuous), results presented in \citet{yeo2013} (dashed), 
 and restored data (diamonds). Black: $\mu$=0.975.
 Red: $\mu$=0.625. Green: $\mu$=0.325. The bar shows the average standard deviation of the values in the bins.    
 \label{fig7}}
\end{figure}

The studies by \citet{ortiz2002} and \citet{yeo2013} were conducted with the MDI and the HMI, respectively and are those that allow a 
more direct comparison with our investigation. 
Figure \ref{fig7} shows 
the contrasts obtained at three angular positions from the cubic fits to our original data (Eq.~\ref{equation}) and those presented in \citet{yeo2013} (their Eq.~3).
For magnetic flux values smaller than 800 G, the curves  present a very good agreement, with differences being within the error of the measurements. 
On the contrary, at larger magnetic flux values the fits presented in \citet{yeo2013} do not seem to represent our measurements, most likely because those authors
did not include high magnetic flux pixels in their analysis (see Sec. \ref{sec:obs}). The agreement is remarkable if we consider the difference in the type of HMI images employed
(but see discussion in Sec.~\ref{sec:disc:hmi})
and the different data reduction strategies employed in the two studies. In particular,
\citet{yeo2013} employed data obtained averaging original HMI 45 s data acquired over a 315-s interval; they also employed a different method to estimate the limb
darkening shape of quiet regions, and different criteria  to select sunspots and pores. 
It is worth to note that \citet{yeo2013} reported the presence of residual patterns in their intensity images compensated for the limb darkening.  
Following a procedure similar to the one described by those authors, we also estimated the residual intensity on our continuum contrast images, and we found an average
 value of about $3\cdot 10^{-5}$, which is about two orders of magnitude smaller than the value reported by \citet{yeo2014},
 so that our data were not compensated for this effect.

The comparison of results obtained by \citet{ortiz2002}
using MDI data with those obtained using HMI are largely discussed in \citet{yeo2013}, and we do not repeat it here.  We only note that, as showed by \citet{liu2012},
 the MDI magnetic flux values are 1.3 - 1.4 times larger that the HMI magnetic flux values, so that HMI contrasts obtained at certain magnetic flux 
 ranges should be compared 
 with MDI contrasts obtained at higher magnetic flux ranges. Nevertheless, as shown by a comparison of results in Fig.~\ref{fig3} and Fig.~\ref{fig4}, with the ones
 reported in Fig.~3 and Fig.~4 of \citet{ortiz2002}, this scaling factor is not sufficient to explain the different contrasts obtained with the two instruments.
 We therefore confirm that the different magnetic flux and angular dependences of the continuum contrast obtained by our analysis and the one by \citet{yeo2013}
 on one side, and
  \citet{ortiz2002} on the other, must be ascribed to the different spatial resolution and the different definition of sunspot regions adopted.

\begin{figure}
\includegraphics[width=8cm,height=6cm,trim=1mm 0 2mm 0mm, clip=true]{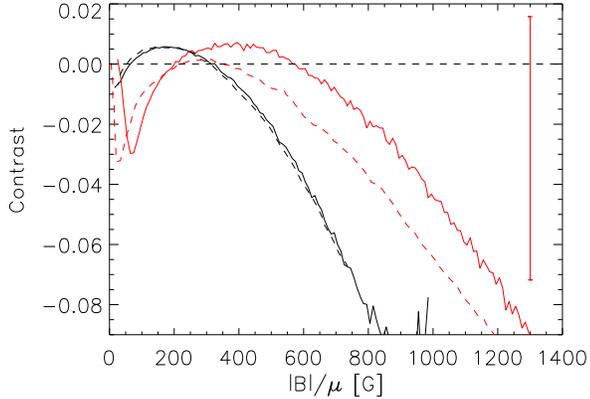}
 \caption{Comparison with results obtained close to disk-center from this study (continuous) and results reported in
 Fig.~15 of \citet{yeo2014} (dashed). Black: original. Red: restored. Data were averaged on 10~G magnetic flux bins. The error bar shows the
 average standard deviation value over the bins obtained for restored data.
 \label{fig8}}
\end{figure}

In Fig.~\ref{fig8} we compare our results with those obtained by  \citet{yeo2014}, who also compensated HMI data for scattered-light effects, 
but employing a different procedure (see Sec.~\ref{sec:obs}), and restricting their analysis of photometric contrast only to disk center.
 The plot shows that the original data produce very similar results.
The restoration produces in both cases an enhancement of the contrast at low magnetic flux values ( $|B|/\mu < $ 200~G), 
and an enhancement of the brightness at higher magnetic flux values, as expected from simulations \citep{rohrbein2011}, but values 
present some discrepancies. 
These differences must be attributed to the limited sample of data employed by \citet{yeo2014} (one day), to the use of one of the filtergrams 
(-344 m$\AA$ from line center) instead of the intensitygram
data-product, to the different criteria employed to 
define sunspots (\citet{yeo2014} employed a -0.11 contrast threshold and a 3-pixel kernel) and, finally, to the different algorithms employed to derive the instrumental PSF. It is difficult to discern which of these effects plays the largest role, as, for instance, the small differences in the original data might have been amplified by the restoration. A detailed comparison of the performance of the two codes employed for the restoration goes beyond the purpose of this paper. Here we note that the differences are overall within the error bars of our measurements.

To the best of our knowledge, the effects of scattered-light compensation on the center-to-limb variation of magnetic features contrast 
cannot be directly compared with any previous studies. Most of the studies performed on high spatial resolution data or on MHD simulations were in fact conducted at or 
close to disk-center \citep[e.g.][]{mathew2009,rohrbein2011, danilovic2013, criscuoli2014b}, while results obtained on full-disk data mostly focused on sunspot properties 
\citep[e.g.][]{walton1999,mathew2007,criscuoli2008} or facular properties but at different wavelength ranges \citep[e.g. CaIIK][]{walton1999,criscuoli2008}. 
Here we note that contrast variations are on average larger for network pixels rather than for facular ones, as a result of the different effects that image degradation 
has on magnetic features of different sizes \citep[e.g.][]{criscuoli2007,criscuoli2008,viticchie2010}.  

\begin{figure}
\includegraphics[width=8cm,height=6cm,trim=10mm 0 2mm 0mm, clip=true]{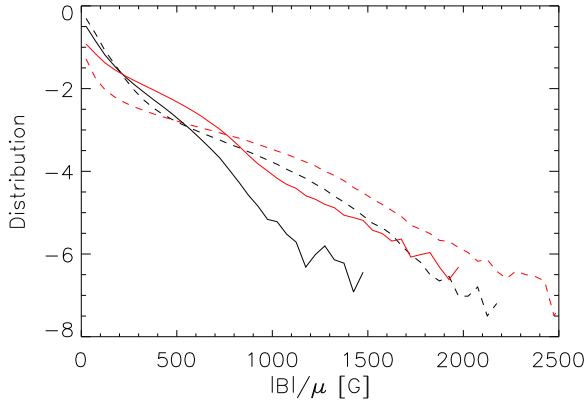}
 \caption{Magnetic flux distributions of pixels located in network (black) and facular (red) regions obtained from original (continuous) and restored data (dashed). 
 \label{fig9}}
\end{figure}

The network and facular contrast differences estimated on restored images are small, especially at disk center, and within the error bars. On the other hand,
these differences are in qualitative agreement with previous studies 
obtained at high spatial resolution close to disk-center (see Sec.~\ref{sec:res}), 
but a quantitative comparison is hampered by the different type of data 
employed.  Our results are also in qualitative agreement with measurements carried out by \citet{ortiz2002}
and \citet{yeo2013}, which showed that the specific contrast of low-magnetic flux pixels is larger than the specific contrast of high-magnetic flux pixels.
On the other hand, our results do not agree with previous observations carried out using full-disk observations, 
which produced a network contrast lower than the facular one, and almost constant over the disk.  As previously suggested in \citet{ortiz2002},
these differences are mostly caused by spatial resolution and scattered-light effects. Indeed, we found that network and facular regions are 
characterized by different contrasts only when using restored data.

\section{Implications for solar irradiance studies}\label{sec:solairra}
In this study we employed spatial proximity criteria to discriminate between network and facular regions. Some previous studies discriminated 
between the two features assuming that statistically
lower magnetic flux 
pixels belong to network, while higher magnetic flux pixels belong to faculae and active network regions. 
In the following we evaluate how these different criteria affect the estimate of magnetic features contrast. 
Figure~\ref{fig9} shows the magnetic field distribution of 
 network and facular regions derived from original and restored data. Original data shows that
for $|B|/\mu <$~200~G pixels are more likely to be located in network than in facular regions. This value is in near-agreement with the 130~G values 
employed by \citet{ortiz2002} and \citet{yeo2013} to distinguish between network and faculae.
The threshold increases to about 400~G when analyzing restored data, and the difference between the 
two populations is up to one order of magnitude. 

Results in Fig.~\ref{fig1A} and in Fig.~\ref{fig5} show that for  $|B|/\mu <$~200~G and $|B|/\mu <$ 400~G, 
respectively, the contrast differences
between network and facular pixels are small, up to approximately 0.02 at the limb, so that the criteria employed in previous studies to separate network from faculae 
allow reasonable 
estimates of network properties. 
At larger magnetic flux values, instead, the contrast differences between network and faculae are larger, especially toward the limb, so that we expect facular contrasts
derived using the sole magnetic flux to discriminate between the two classes of features to be statistically affected by the network. 

\begin{figure}
\includegraphics[width=8cm,trim=1 5mm 2mm 2mm, clip=true]{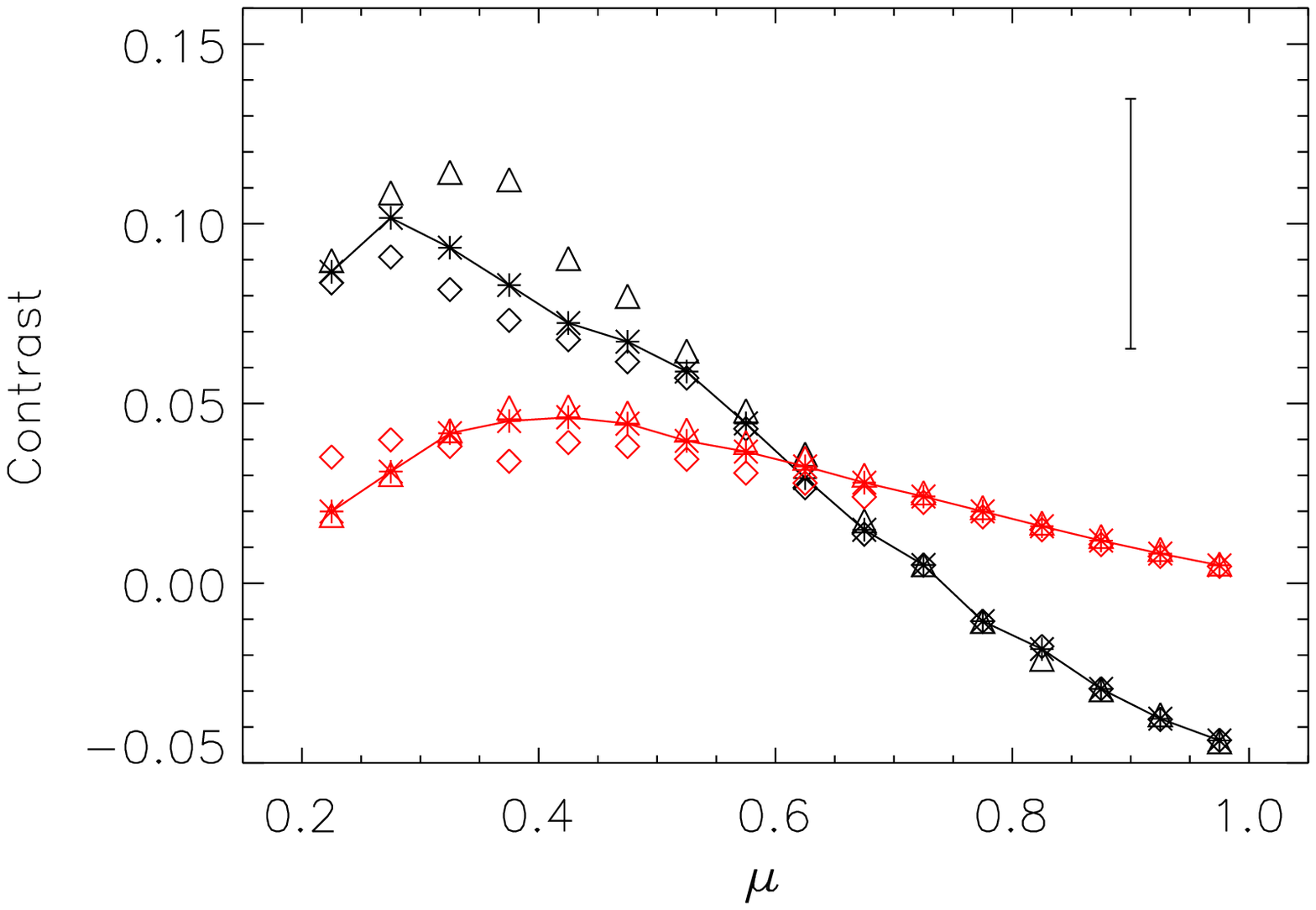}\\
\includegraphics[width=8cm,trim=1 4 2mm 3mm, clip=true]{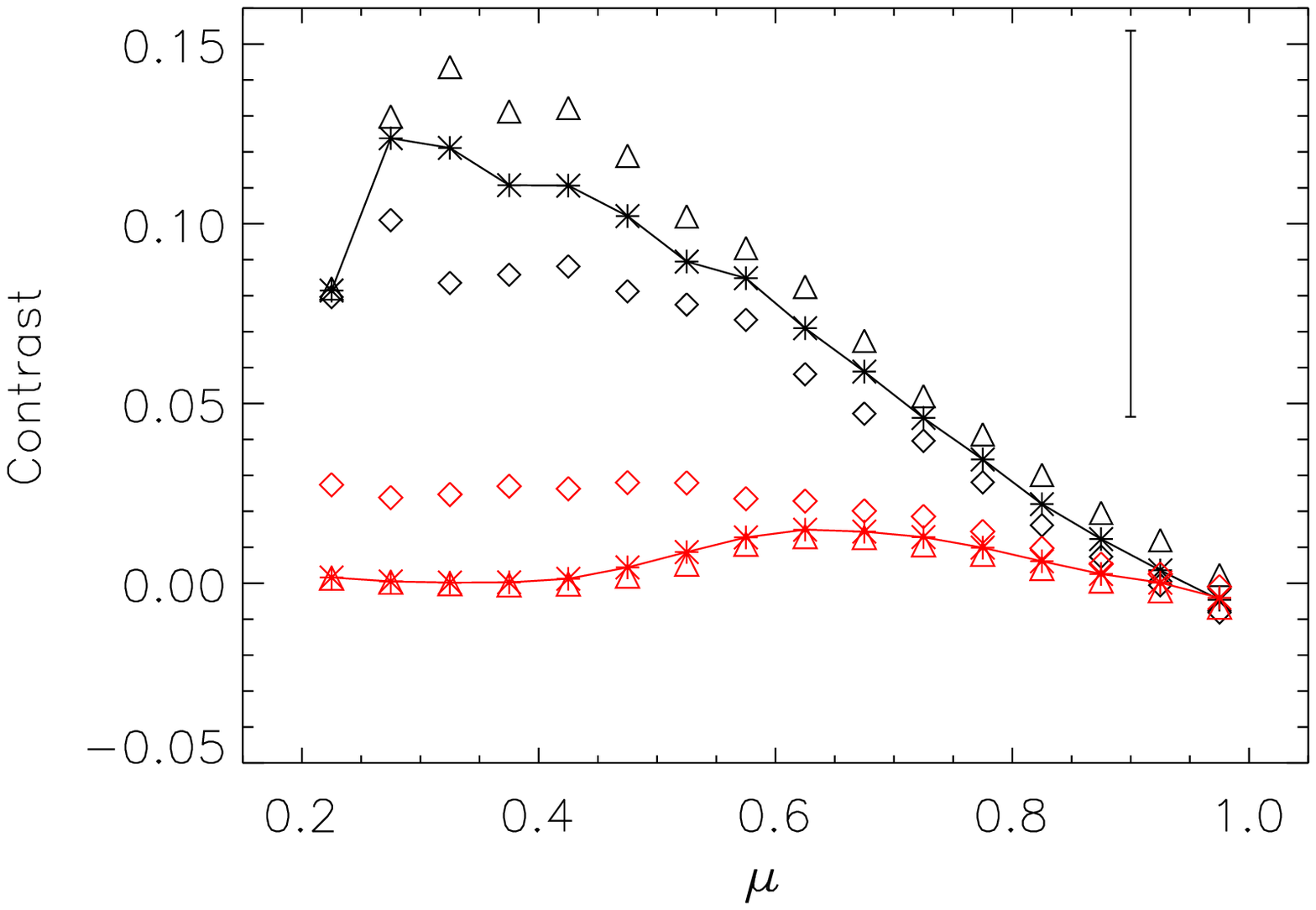}
 \caption{Center-to-limb variation of the continuum contrast derived from original (top) and restored (bottom) data. 
 Black symbols correspond to pixels where 600~G$<|B|/\mu <$ 650~G and red to pixels where 200 G$<|B|/\mu <$ 250~G. Triangles denote network pixels,
 diamonds facular pixels
 and stars-continuous lines denotes results obtained without discrimination. The error bar corresponds to 
 the typical standard deviation over the bins. 
 \label{fig10}}
\end{figure}

This is confirmed by results in
Fig.~\ref{fig10}, which show as an example the contrasts of network and facular pixels with magnetic fluxes of $\approx$ 200~G (red lines) and $\approx$~600~G (black lines), 
together with the contrast 
derived without distinguishing between the two features, obtained from original (top) and restored (bottom) data. Both panels in Fig.~\ref{fig10} show that the contrast derived for 
$|B|/\mu \approx$~200~G without distinguishing between network and faculae, closely follow the results obtained for the network. For $|B|/\mu \approx$~600~G, instead,
the contrast derived without distinguishing between the two features is in between the values found for network and faculae, thus showing that the network 
statistically affects the estimates of facular contrast if faculae are singled out only according to the magnetic flux value. The effects are larger 
toward the limb and on restored images, where the network and facular contrast differences are larger.
\\

To evaluate the effects of discriminating between network and faculae on the estimate of solar irradiance variations over the solar activity cycle, we 
computed the daily facular and network contribution to TSI \citep[c.f.r.][]{lean1998,foukal1991}:

\begin{eqnarray}
\label{eqdeltaF}
 \frac{\Delta F}{F} & = & \sum\limits_{k} \sum\limits_{j} \frac{5 \mu_j N(\mu_j,B_k)C(\mu_j,B_k)\Psi(\mu)}{2}
\end{eqnarray}

where the two sums run over the magnetic flux and the cosine of the heliocentric angle, $N(\mu_j,B_k)$ is the area of pixels at position $\mu_j$ 
and magnetic flux $B_k$
normalized to the surface of the solar hemisphere, $C(\mu_j,B_k)$ is the contrast as derived by the bi-cubic fits to our data and 
$\Psi(\mu)=(3\mu+2)/5$ is the quiet Sun
limb-darkening function.  It is important to note that this model is raher simple, especially if compared to modern irradiance reconstruction techniques,
as it lacks of detailed knowledge of the radiometric contribution of magnetic features to the disk integrated irradiance.  In particular, because the
bolometric contrast is knwon to be larger than the one measured in the red continuum \citep{foukal2004}, our computations 
underestimate the bolometric facular/network contribution to TSI.

The daily facular/network coverage $N(\mu_j,B_k)$ was estimated employing original 45 s HMI data acquired between April 2010 and October 2015.
Coherently with the analysis presented above,
network and facular regions were discriminated using the HARP masks.  The facular excess was then computed using Eq.~\ref{eqdeltaF} first employing the $C(\mu_j,B_k)$
curves derived from the whole dataset (Model A, in the following), second using the contrast curves derived for network and faculae separately (Model B, in the following). 
The minimum value of the magnetic flux considered is 300 G. This value was chosen for two reasons. First, the bi-cubic fits seem to reproduce best 
the observations
for magnetic flux values above this threshold (for smaller magnetic flux values, toward the limb the fits overestimate the contrast up to 0.03). 
Second, as discussed in \citet{schnerr2011}, at magnetic flux values smaller than this threshold 
the contrast of small, unresolved magnetic features is largely underestimated because of the brightness contribution of the dark lanes they are embedded in.

\begin{figure}
\includegraphics[width=9cm,trim=6mm 0 5mm 0mm, clip=true]{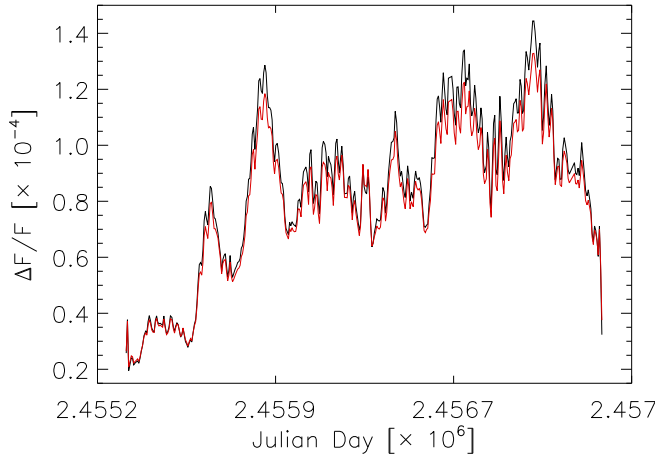}
 \caption{Temporal variation of the network and facular contribution to irradiance for Model A (black) and Model B (red). 
 \label{fig11}}
\end{figure}

Results presented in Fig.~\ref{fig11} show that Model A overestimates the TSI excess, mostly because, as shown in Fig.~\ref{fig10}, the facular contribution
during the periods of high activity is overestimated. The variability measured between the periods of 
largest and lowest activity ($\Delta F/F _{Max} - \Delta F/F_{Min}$ ) is $1.25\cdot 10^{-4}$ and $1.12\cdot 10^{-4}$ for Model A and Model B, respectively,
which corresponds to a difference 
of about 11\%. This estimate must be considered a lower limit for several reasons.  The most important, is that in our analysis we discarded pixels with magnetic flux smaller than 300~G, which 
occupy the majority (more than 98\%) of the solar magnetized surface. Indeed, \citet{schnerr2011}  employed high-spatial resolution data acquired in a red continuum to show
that the contribution of magnetic pixels with flux larger than 300 G to the network brightness is only $0.2\cdot 10^{-4}$, 
compared to the excess brightness of about $1.1\cdot 10^{-3}$ found when including all magnetized network pixels. We note that $0.2\cdot 10^{-4}$ is
in reasonable agreement with the
$0.16\cdot 10^{-4}$ value that we found for the network during 2010 (not shown), the period of lowest magnetic activity in our observations.
Spatial resolution effects, 
(which reduced the estimate of contrasts, magnetic fluxes, and the contrast difference between network and faculae) and the fact that our observations did not include
a period of minimum (the HMI operations started during the rising phase of cycle 24), also contribute to underestimate both
$\Delta F/F$ and the differences between Modelels A and B. Finally, the use of original data instead of restored ones to derive the coverage of magnetic pixels 
is another source of uncertainty, as the restoration changes the distribution of magnetic pixels. In particular, Fig.~\ref{fig9} shows that 
at magnetic flux values larger than approximately 500~G the difference between the number of facular and network pixels is smaller in restored than in original data, 
which might have lead to an overestimate of the variability difference produced by the two models.

It is also important to note that the facular/network contributions to irradiance variations were estimated using contrasts averaged in magnetic field bins.  
Our results show that the restoration largely increases the scatter of the contrast in each bin, thus pointing
to the necessity of 
employing 
criteria to discriminate between various types of magnetic features other than the magnetic flux alone. In particular, as already noted in Sec.~\ref{sec:obs},
part of the pixels with $|B|/\mu >$ 800~G 
on restored images present a negative contrast at all angular positions. Although on average the brightness increases after restoration, the fraction of negative contrast
pixels at these magnetic flux ranges is more than 50\%,
thus suggesting a different temperature stratification than pixels characterized by a positive contrast. If, as suggested in \citet{foukal2015} \citep[see also][]
{foukal1993,foukal1990}, the coverage of such features increases with the activity, then 
available reconstruction techniques making use of facular 
coverage as derived by chromospheric emission \citep[e.g.][]{lean2000, foukal2012}
overestimate the contribution of faculae to irradiance, especially during the strongest cycles. A detailed study of properties of dark faculae is under investigation.

Finally, it is important to mention that most recent irradiance reconstruction models do not make use of direct measurement of 
photometric contrast (if not in some cases 
as a proxy e.g. \citealt{chapman2012}) and that we expect the amount of uncertainty introduced by not taking into account network and faculae separately 
to vary for different techniques. Nonetheless, we note that the uncertainty value of $\approx$ 10\% in the contribution of network and faculae to TSI 
variations
is smaller than or of the same order of uncertainties reported
for some TSI and SSI reconstructions \citep[e.g.][]{crouch2008,yeo2014c,coddington2015}. 

\section{Conclusions}\label{sec:conc}

We employed HMI data-products compensated for the instrumental PSF to compare the center-to-limb variation of the contrast of network and 
facular regions as determined by their proximity to active regions in addition to continuum intensity and magnetogram threshold.
Contrary to previous results obtained on full-disk observations, we found that for magnetic flux values above 300~G the network is brighter than faculae 
and presents a stronger contrast center-to-limb variation.
This finding for full-disk data must be attributed to the increased photometric contrast resulting from correcting for scattered-light. Note that the 300~G thresold 
is function of the spatial-resolution of the employed data (see also discussion in Sec.~\ref{sec:solairra}). Our results are in
qualitative agreement with those obtained at or close to disk-center by the analysis of sub-arcsecond observations and MHD simulations. 
 We extend the analysis to full-disk  
to report the photometric contrasts of network and faculae, as determined based on proximity to active regions, as a function of magnetic flux and 
line-of-sight observing angle.

We employed a simple model to estimate the contribution of magnetic pixels to TSI variations 
($\Delta TSI$) and found that if the contribution of network and faculae is not taken
into account separately, or is only based on the magnetic flux of a pixel, $\Delta TSI$ is overestimated of about 11\%. 
This value is a lower bound for error for the following reasons. First, the contrast measured using HMI images is still affected by limited spatial resolution.
Second, in our analysis we discarded pixels located in dark lanes, so that the contribution of unresolved 
bright structures located in these regions is not taken into account. Third, comparison of results obtained on original and restored data shows that for each
magnetic flux bin the restoration increases the scatter of results, even when taking into account faculae and network separately. This suggests that classification 
of features according to the magnetic flux value and spatial aggregation might not be enough to properly characterize the contribution of magnetic elements to
irradiance variations, if irradiance reconstruction techniques just employ the above parameters as input data.

Our report on the photometric contrasts of faculae and network, defined using spatial proximity to active regions, as a function of
center-to-limb angles can be used to further improve existing irradiance reconstruction techniques. A majority of irradiance reconstruction
techniques either do not explicitly differentiate between features, or assume a lower contrast for the network based on previous studies.
The continuity and frequency of full-disk HMI data, combined with the ease by which HARP masks allow identification of active region locations,
plus the availability of a fast routine for removal of scattered-light, means that a daily, corrected HMI images of similar quality could easily
be implemented for irradiance modeling. 

Finally, our results support the conclusion of \citet{foukal2011} that the contribution of network during the Maunder and Sp\"{o}rer minima might be
underestimated by irradiance reconstruction models \citep[e.g.][]{crouch2008,tapping2007,wang2005,lean2000}, 
that assume a linear relation
between magnetic flux and/or plage coverage and irradiance.

\acknowledgments
This work was
carried out through the National Solar Observatory Research
Experiences for Undergraduate (REU) site program, which is co-funded by the
Department of Defense in partnership with the NSF REU Program. The National
Solar Observatory is operated by the Association of Universities for Research in
Astronomy, Inc. (AURA) under cooperative agreement with the National Science
Foundation. S.C. is grateful to Dr. Peter Foukal for reading the paper and providing useful comments, and to Dr. Odele Coddington for the interesting discussions 
about the NRL reconstructions.

\bibliography{thebib}



\appendix

\section{Comparison of Quiet and Active regions derived from original data}\label{A1}

\begin{figure*}[!h]
\includegraphics[width=16cm,height=10.6cm, trim=3mm 4 1mm 4mm, clip=true]{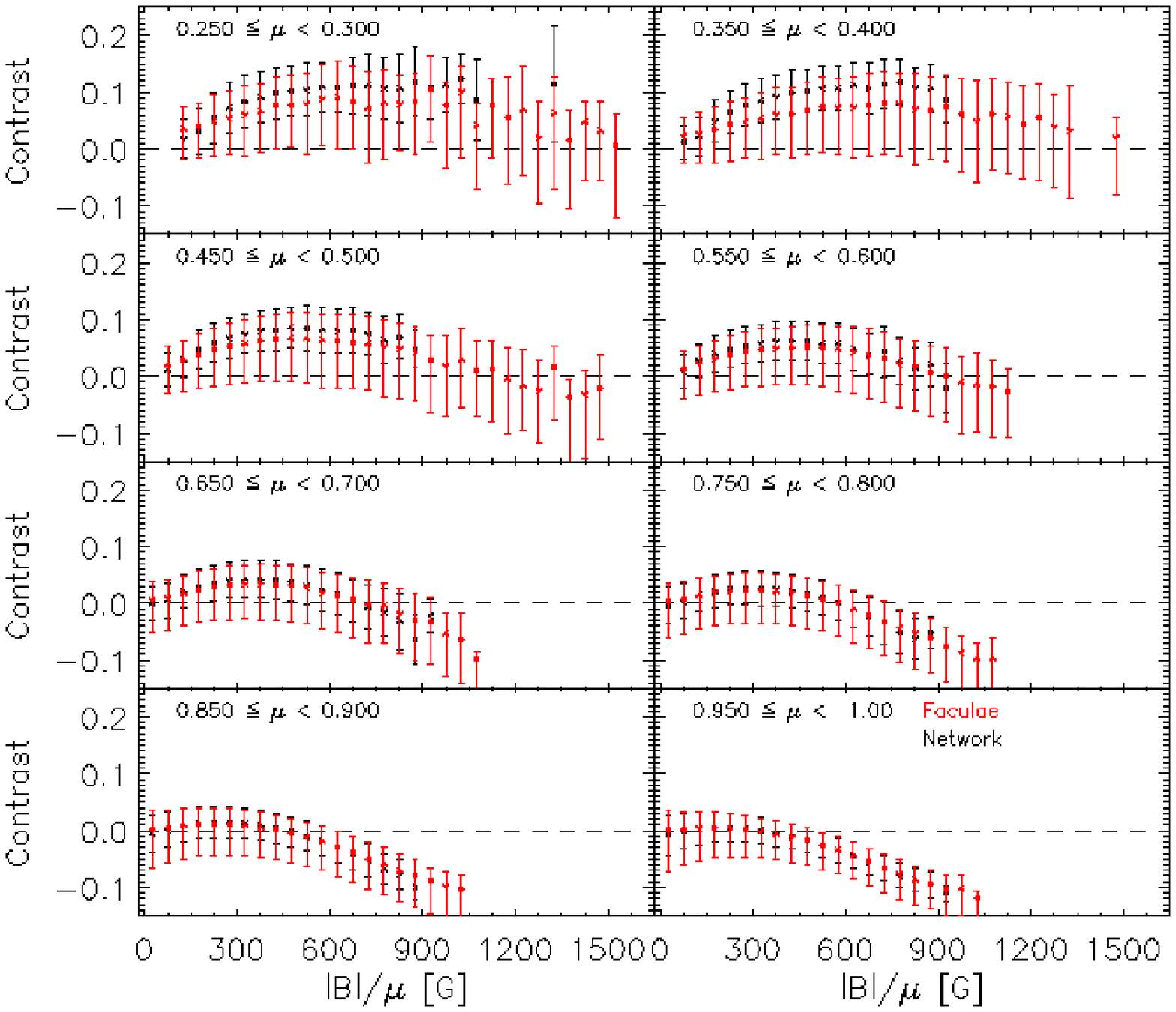}
 \caption{Variation of the intensity contrast with the magnetic flux for pixels located at various radial distances from disk-center in facular (red)
 and in network (black) regions singled out on original data.
 \label{fig1A}}
\end{figure*}

\begin{figure*}[!h]
\includegraphics[width=16cm,height=10.6cm, trim=3mm 4 1mm 4mm, clip=true]{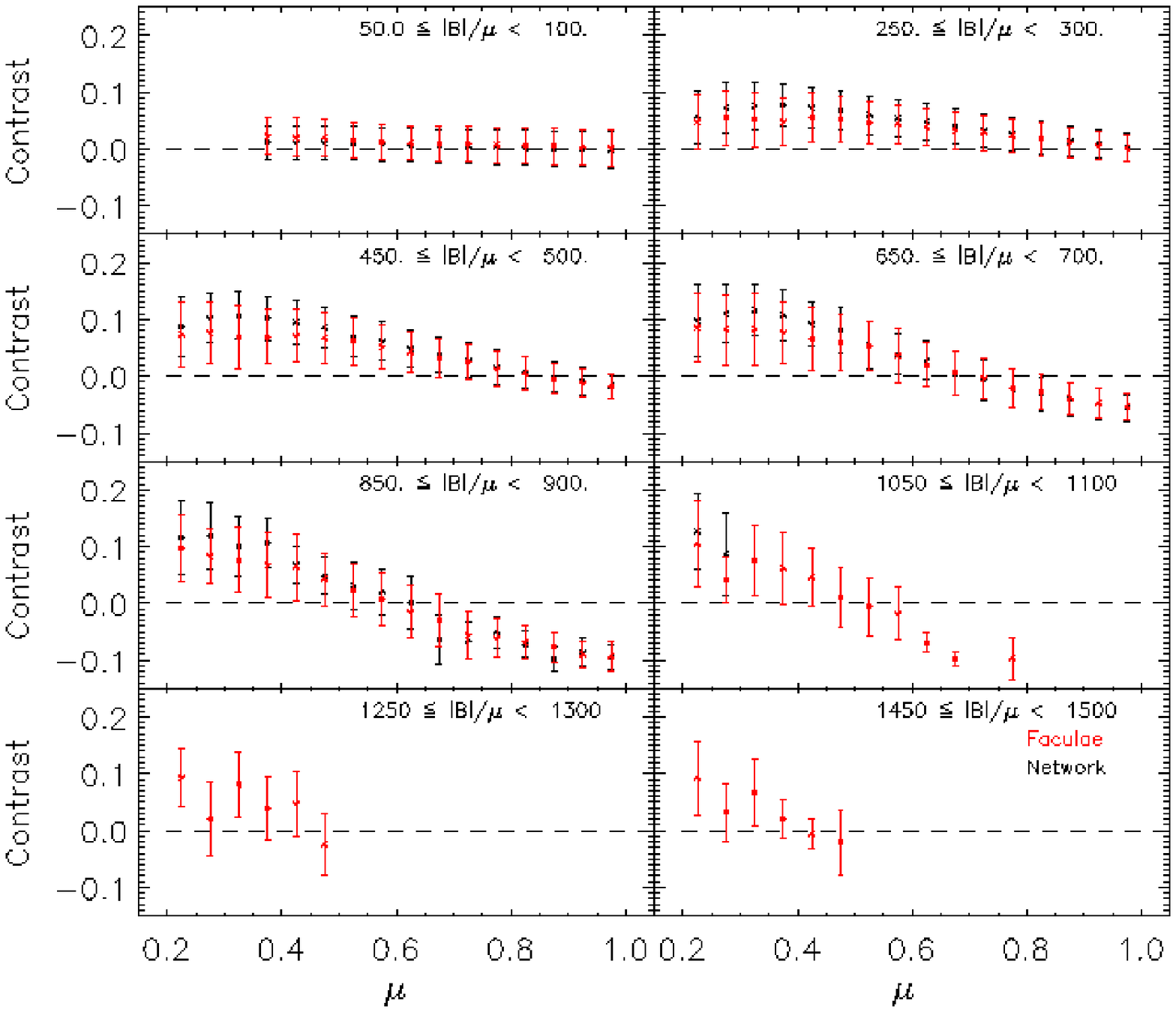}
 \caption{Variation of the intensity contrast with the cosine of the heliocentric angle in
 facular (red) and in network (black) regions singled out on original data. 
 \label{fig2A}}
\end{figure*}

\clearpage

\section{Polynomial surface fit coefficients}\label{A2}

The analytical formula is the following:  
\begin{eqnarray*}
\label{equation}
C\left(\mu,\frac{B}{\mu}\right)
 & = & 
\left[
\begin{array}{c}
 
10^{-2} \left(\frac{B}{\mu}\right)^0 \\
10^{-3} \left(\frac{B}{\mu}\right)^1 \\
10^{-6} \left(\frac{B}{\mu}\right)^2 \\
10^{-9} \left(\frac{B}{\mu}\right)^3 \\

\end{array}
\right]^T  
 \left[ \mathcal{M}
\right]  
 \left[
\begin{array}{c}
 \mu^{0} \\
 \mu^{1} \\
  \mu^{2} \\
   \mu^{3} \\
\end{array}     \right]  
\end{eqnarray*}

Fit coefficients original-data:
\begin{equation}
 \mathcal{M} = \left[ \begin{array}{cccc}
-6.77 &      28 &     -43 &    21\\
   0.73  &   -1.46 &    2.07 &    -1.18\\
   -1.22 &   4.37 &  -7.99 &   4.35\\
    0.73 &  -3.31 &   5.72 &  -2.91
\end{array} \right]  
\end{equation}

Fit coefficients for deconvolved-data:
\begin{equation}
 \mathcal{M} = \left[ \begin{array}{cccc}
      -16.62 &      44. &    -54.6 &      24 \\
   0.47 &   0.71 &   -1.55 &   0.56 \\
  -0.25 &  -1.02 &   1.05 & -0.083\\
   0.082 &   0.085 &   0.095 &  -0.16

\end{array} \right]  
\end{equation}

Fit coefficients for Network deconvolved-data:
\begin{equation}
 \mathcal{M} = \left[ \begin{array}{cccc}
      -15.068    &   33.757 &     -36.765 &    14.03\\
  -00.289  &  4.333 &   -6.918 &    3.115 \\
 1.452 & -8.771 &   12.349 &  -5.373\\
  -0.739 &   3.9512 &  -5.670 &   2.570

\end{array} \right]  
\end{equation}

Fit coefficients for Faculae deconvolved-data:
\begin{equation}
 \mathcal{M} = \left[ \begin{array}{cccc}
      -0.948 &     2.56 &     -10.96 &     7.41 \\
    0.239 &   0.184 &   0.044 &  -0.346\\
 -0.7036 &   2.747 &  -5.168 &  2.919\\
  0.5589 & -2.601 &   4.116 &  -2.020

\end{array} \right]  
\end{equation}
\pagebreak

\section{10-th order polynomial fit coefficients}\label{A3}
The magnetic flux dependence of the contrast was fitted separately for different positions over the disk $\mu$:
\begin{equation}
 C\left(\frac{B}{\mu}\right)=\sum \limits_{k=0}^{k=10} a_k \cdot \left(\frac{B}{\mu}\right )^k
\end{equation}

The coefficients of the fits are given for different $\mu$ intervals in Table \ref{Tab2}

\begin{table}[h]
 \caption{Coefficients derived fitting the restored data with a 10-th order polynomial. } 
  \label{Tab2}
 \resizebox{1.05\textwidth}{!}{  
 \begin{tabular}{c|ccccccccccc}

\hline
$\mu$ & $a_0\cdot10^{-2}$ & $a_1\cdot10^{-3}$ & $a_2\cdot 10^{-5}$ & $a_3\cdot 10^{-7}$ & $a_4\cdot 10^{-10}$ & $a_5\cdot 10^{-12}$ & $a_6\cdot 10^{-15}$ & $a_7\cdot 10^{-18}$ & $a_8\cdot 10^{-21}$ & $a_9\cdot 10^{-25}$ & $a_{10}\cdot 10^{-29}$ \\
\hline
          
$1 - 0.95$  & 5.14 &  -2.86 & 3.89 & -2.50 & 9.21 & -2.11 &  3.12 & -2.95 & 1.73 & -5.76 &  8.24\\
$0.95 -  0.9$  & 5.02 &  -2.62 & 3.47 & -2.14 & 7.61 & 2.38 &  -2.17 & 1.22 & -3.91 & 5.37 &  8.24\\ 
$0.9 - 0.85$  & 7.07 &  3.31 & 4.39 & -2.78 & 1.02 & -2.33 &  3.43 & -3.25 & 1.91 & -6.31 &  9.00\\ 
$0.85 - 0.8$  & 7.10 &  3.08 & -3.86 & -2.29 & 7.89 & -1.71 &  2.41 & -2.21 & 1.26 & -4.11 &  5.81\\ 
$0.8 - 0.75$  & 11.35 &  4.51 & 5.72 & -3.51 & 1.26 & -2.84 &  4.15 & -3.92 & 2.30 & -7.68 &  11.06\\
$0.75 - 0.7$  & 10.32 &  -3.91 & 4.69 & -2.68 & 8.89 & -1.85 &  2.48 & -2.15 & 1.16 & -3.57 & 4.76\\
$0.7 - 0.65$  & 11.57 &  -4.26 & 5.14 & -3.00 & 1.04 & -2.28 &  3.29 & -3.08 & 1.82 & -6.10 & 8.905\\
$0.65 - 0.6$  & 15.28 &  -5.14 & 5.94 & -3.37 & 1.14 & -2.45 &  3.47 & -3.20 & 1.86 & -6.14 & 8.82\\
$0.6 - 0.55$  & 6.46 &  -1.69 & 0.98 & 0.21 & -3.42 & -1.30 &  -2.58 & 2.98 & -2.04 & 7.64 & -12.06\\
$0.55 - 0.5$  &9.57& -2.43&  1.77& -0.32& -1.11&  0.66 & -1.45&  1.76& -1.23&  4.66 & -7.41\\
$0.5 - 0.45$  &0.59&   0.96&  -2.84&   2.78& -12.94& 3.44&-5.59 &  5.69&  -3.53 & 12.20& -18.05\\
$0.45 - 0.4$  &-14.60& 5.68& -8.32&   5.98& -23.80&   5.75&  -8.78&   8.56&  -5.15&  17.44& -25.43\\
$0.4 - 0.35$  &-21.77&   6.68& -7.96& 4.77& -16.02& 3.27&-4.22 &  3.44&  -1.72 &  4.78 & -5.60\\
$0.35 - 0.3$  &-39.17&  10.40& -10.84&   5.83& -18.13&   3.52&  -4.42&   3.60&  -1.84&   5.35& -6.80\\
$0.3 - 0.25$  &16.61&  -4.78 &  5.80 & -3.79 & 14.56&  -3.42  & 5.07&  -4.75&   2.74 & -8.89 & 12.39\\
$0.25 - 0.2$  &151.39& -33.19&  30.59 &-15.60&  48.81&  -9.84&  13.05& -11.30 &  6.16& -19.19 & 26.03\\
\end{tabular}}
\end{table}

\end{document}